\documentclass{aastex631}

\shorttitle{High Energy Emission of PKS 1502+036}
\shortauthors{Wang et al.}

\graphicspath{{./}{figures/}}
\usepackage{appendix}
\usepackage{amsmath}
\usepackage{color}
\begin{document}

\title{On the Hadronic Origin of High Energy Emission of $\gamma$-ray Loud Narrow-Line Seyfert 1 PKS 1502+036}

\author{Zhen-Jie Wang}
\affiliation{Department of Astronomy, Xiamen University, Xiamen, Fujian 361005, China}

\author{Ze-Rui Wang}
\affiliation{College of Physics and Electronic Engineering, Qilu Normal University, Jinan 250200, China}

\author[0000-0003-1576-0961]{Ruo-Yu Liu}
\affiliation{School of Astronomy and Space Science, Xianlin Road 163, Nanjing University, Nanjing 210023, China}
\affiliation{Key laboratory of Modern Astronomy and Astrophysics (Nanjing University), Ministry of Education, Nanjing 210023, China}

\author[0000-0003-4874-0369]{Junfeng Wang}
\affiliation{Department of Astronomy, Xiamen University, Xiamen, Fujian 361005, China}

\newcommand{\ry}[1]{{\color{red} #1}}
\correspondingauthor{Ze-Rui Wang; Junfeng Wang}
\email{ zeruiwang@qlnu.edu.cn; jfwang@xmu.edu.cn}
\begin{abstract}

    The radiation mechanism of Radio-Loud Narrow-Line Seyfert 1 (RL-NLS1) from X-ray to $\gamma$-ray bands remains an open question. While the leptonic model has been employed to explain the spectral energy distribution (SED), the hadronic process may potentially account for the high energy radiation of some $\gamma$-ray loud Narrow-Line Seyfert 1 (NLS1) as well. We study one of such RL-NLS1, PKS 1502+036, comparing the theoretical SEDs predicted by the leptonic model and the lepto-hadronic model to the observed one. For the hadronic processes, we take into account the proton synchrotron radiation and proton-photon interactions (including the Bethe-Heitler process and the photopion process) including the emission of pairs generated in the electromagnetic cascade initiated by these processes. Our results show that the leptonic model can reproduce the SED of this source, in which the X-ray to $\gamma$-ray radiation can be interpreted as the inverse Compton (IC) scattering. On the other hand, the proton synchrotron radiation can also explain the high energy component of SED although extreme parameters are needed. We also demonstrate that the $p\gamma$ interactions as well as the cascade process cannot explain SED. Our results imply that a leptonic origin is favored for the multi-wavelength emission of PKS~1502+036.
    

\end{abstract}


\section{Introduction} \label{sec:intro}
    
    Blazars are a class of active galactic nuclei (AGNs) with relativistic jets pointing towards the observer and known to exhibit high amplitude $\gamma$-ray flux variations \citep{1995PASP..107..803U}. They are widely speculated to be strong cosmic-ray proton (or nuclei) accelerators and have long been considered as potential emitters of high-energy electromagnetic (EM) radiation and neutrino radiation through the hadronic interactions of these accelerated cosmic rays \citep[e.g.,][]{1993A&A...269...67M, 1995APh.....3..295M, 1996SSRv...75..341S, 2001A&A...367..809K, Halzen_2002, 2001PhRvL..87v1102A,  2014PhRvD..90b3007M}. Such speculations are supported by the discovery of possible correlation between high-energy neutrino events and blazars \citep{2018Sci...361.1378I, 2018Sci...361..147I}.
    These events have been extensively studied in the framework of the so-called photohadronic interaction model \citep{Ansoldi_2018, 2018ApJ...863L..10A, 2018ApJ...864...84K, 2018ApJ...865..124M,     2018MNRAS.480..192P, 2019MNRAS.484L.104P,  2019MNRAS.483L..12C, 2019NatAs...3...88G, 2019ApJ...881...46R, 2019ApJ...874L..29R, Xue_2019, IceCube:2021oqh} and hadronuclear interaction model \citep{2018ApJ...866..109S, Liu_2019, 2019PhRvD..99j3006B}. As such, hadronic process seems to be widely used to explain the broadband SED of jets in blazars.

    In blazars, the broadband SED of jets have been broadly studied \citep{1995A&AS..109..267G, 1995ApJ...440..525V, 1996ApJ...463..444S, 1998MNRAS.299..433F, 2006A&A...445..441N, 2010ApJ...716...30A}, but the high energy radiation mechanism of jet in AGNs is still controversial.
    Similar to blazars, RL-NLS1 objects are also a class of AGN hosting highly relativistic non-thermal jets \citep{Doi_2006}, characterized by narrow Balmer lines (FWHM $\rm H_{\beta}\ <\ 2000\ km\ s ^{-1}$), weak [O III], and strong Fe II emission \citep{Osterbrock_1985}. The detection of $\gamma$-ray emissions from RL-NLS1s by Fermi/LAT is believed to be an evidence for the existence of aligned relativistic jets in this class of AGNs \citep{Paliya_2018,Paliya_2019,Zhang_2020}. It is rather plausible that relativistic protons may exist in their jets and have contribution to the observed radiation. Observations of the first blazar neutrino candidate, TXS 0506+056, favor a mixed scenario with a leptonic dominated SED with subdominant hadronic components in the form of pair-cascades emerging in the hard-X-rays and the TeV band \citep{Cerruti_2020}.
    
    There are sixteen NLS1 galaxies known in $\gamma$-rays \citep{Paliya_2019}, and the high energy radiation mechanism of jet in this type of sources is under debate. The seminal work by \citet{Paliya_2019} presented the results of a detailed multi-wavelength study of this sample of sixteen NLS1 galaxies known in $\gamma$-rays so far, and successfully used the leptonic radiative processes to reproduce the broadband SED. Nevertheless, the hadronic radiative processes were not yet considered, which could possibly reproduce the broadband emission as well. It is worth exploring whether the hadronic model can explain the high energy emission of jets in NLS1s, which became the motivation of this work. 
   
     The RL-NLS1 galaxy PKS 1502+036 was found to be emitting in $\gamma$-ray band by \emph{Fermi}/LAT \citep{Orienti_2012,Paliya_2018}, and was a promising neutrino source candidate in the catalog of \citet{Aartsen_2020}, with a local pre-trial $p$-value 0.28 \citep{Aartsen_2020}. PKS 1502+036 has rich multi-band observation data from radio to $\gamma$-ray band, so we select PKS 1502+036 as the prime target to study in this work. Located at $z=0.409$ \citep{2009ApJ...707L.142A_Abdo}, it is a faint but persistent $\gamma$-ray emitter \citep{2015ApJ...808L..Paliya}. The radio morphology shows a core-jet structure from Very Long Baseline Array imaging\citep{Orienti_2012}.  The broadband SED of PKS 1502+036 can be described using the synchrotron radiation of an electron population and inverse Compton scattering of the broad line region (BLR) \citep{Paliya_2019}. 
    
    
    In this work, we employ leptonic and lepto-hadronic models to explain the SED from radio to $\gamma$-ray band of the jet in PKS~1502+036, and investigate the high-energy radiation mechanism and properties of the jet. The Optical-Ultraviolet data and the X-ray to $\gamma$-ray data are taken from \citet{Paliya_2019}, the rest of the data are from NASA/IPAC Extragalactic Database (NED\footnote{\url{https://ned.ipac.caltech.edu/}}), the upper limit on neutrino observation is taken from \citet{IceCube_2020}. The method for calculating the radiation is described in Section 2. In Section 3 we show the SED modeling and results. The maximum injection luminosity of proton under the leptonic model is provided in Section 4. Finally we present discussion and summarize our findings in Section 5. Throughout the paper, $H_0=$71 km s$^{-1}$ Mpc$^{-1}$, $\Omega_{m}=0.27$, and $\Omega_{\Lambda}=0.73$ are adopted.

\section{METHOD}

    To study the high energy radiation mechanism of PKS 1502+036, we use the one-zone leptonic model and the one-zone lepto-hadronic model to reproduce the broadband SED. Similar to the standard radiation model for the blazar jet, we consider the electron synchrotron radiation, the inverse Compton scattering on the synchrotron radiation field (i.e., synchrotron self-Compton, SSC, \citealt{Harris_2006}) and on the external radiation (i.e., extenral Compton, EC) of electrons. For the lepto-hadronic model, we take into account the proton synchrotron radiation, the Bethe-Heitler process, and the photopion process, as well as the emission of pairs generated in the electromagnetic cascade initiated by these processes. It is assumed that relativistic particles are injected into a blob which has a spherical geometry with a radius $R$, filled with a uniformly entangled magnetic field $B$. In the leptonic model, we assumed that the jet moves with a bulk Lorentz factor $\Gamma$, then we have $\delta = 1/[\Gamma - \left(\Gamma^{2} - 1\right)^{1/2}\cos\theta]$ for a relativistic jet in PKS 1502+036 with a viewing angle of $\theta$, where the viewing angle is assumed to be 3$^{\circ}$ \citep{Paliya_2019}. In the lepto-hadronic model, we adopt $\delta$ = 6.6 \citep[the variability Doppler factor in ][]{D_Ammando_2013}.

    Relativistic electrons or protons are usually assumed to be injected in the blob with a broken power-law distribution \citep{2010MNRAS.402..497G,2020_Zhenjie} or a power-law distribution, i.e.,
    \begin{equation}
    \begin{aligned}
    Q_{\rm e(p)}(\gamma_{\rm e(p)})=\ Q_{\rm e(p),0}\gamma_{\rm e(p)}^{-n_{\rm e(p),1}}\left[1+\left(\frac{\gamma_{\rm e(p)}}{\gamma_{\rm e(p),b}}\right)^{\left(n_{\rm e(p),2}-n_{\rm e(p),1}\right)}\right]^{-1},
    \end{aligned}
    \end{equation}
    \begin{equation}
    \begin{aligned}
    Q_{\rm e(p)}(\gamma_{\rm e(p)})&=\ Q_{\rm e(p),0}\gamma_{\rm e(p)}^{-n_{\rm e(p)}},
    \end{aligned}
    \end{equation}
    Here $\gamma_{\rm e(p)min}\ \textless\ \gamma_{\rm e(p)}\ \textless\ \gamma_{\rm e(p)max}$, where $Q_{\rm e(p),0}$ is the normalization, $\gamma_{\rm e(p),b}$ is the break Lorentz factor, $n_{\rm e(p),1}$ and $n_{\rm e(p),2}$ represent the spectral indices below and above $\gamma_{\rm e(p),b}$, $\gamma_{\rm e(p)min}$ and $\gamma_{\rm e(p)min}$ are the minimum and maximum electron or protons Lorentz factors. For electrons, after giving an electron injection luminosity, $Q_{\rm e,0}$ can be obtained from $\int Q_{\rm e}\gamma_{\rm e}m_{\rm e}c^{2}d\gamma_{\rm e} = L_{\rm e,inj}/\left(4/3\pi R^{3}\right)$ where $c$ is the speed of light and $m_{\rm e}$
    is the electron rest mass, $L_{\rm e,inj}$ is the injection luminosity of electrons. For protons, $Q_{\rm p,0}$ can be obtained from $\int Q_{\rm p}\gamma_{\rm p}m_{\rm p}c^{2}d\gamma_{\rm p} = L_{\rm p,inj}/\left(4/3\pi R^{3}\right)$ where $m_{\rm p}$ is the proton rest mass, $L_{\rm p,inj}$ is the injection luminosity of protons. The steady-state electron(protons) distribution can be approximated as
    \begin{equation}
    N_{\rm e(p)}(\gamma_{\rm e(p)}) = Q_{\rm e(p)}(\gamma_{\rm e(p)})t_{\rm e(p)},
    \end{equation}
    where $t_{\rm e(p)} = $\rm min$ \left\{t_{\rm cool},t_{\rm e(p),dyn}\right\}$. For electrons $t_{\rm cool} = 3m_{\rm e}c / \left({4\left(U_{\rm B}+\kappa_{\rm KN}U_{\rm ph}\right)\sigma_{\rm T}\gamma_{\rm e}}\right)$ is the electron radiative cooling timescale, with $U_{\rm B}$ being the energy density of the magnetic field, and $U_{\rm B} = B^{2}/8\pi$. $U_{\rm ph}$ being the energy density of the soft photons. $\sigma_{T}$ is the Thomson scattering cross section and $\kappa_{\rm KN}$ is a numerical factor accounting for the Klein-Nishina effect\citep{2005MNRAS.363..954M}, $t_{\rm e(p),dyn} = R/c$ is the the dynamical timescale of electrons or protons, here $R = r\sin \theta$ is the radius of the blob, $r$ is the distance between black hole and the blob. All timescales are evaluated in the comoving frame of the blob.
    
    In the leptonic model, the synchrotron, SSC and EC radiations are calculated following \citet{Wang_2022}. For the EC process, we consider the radiation of the broad line region (BLR) and the dust torus (DT) as target photons. The BLR and DT radiation is taken as an isotropic blackbody with a peak at $\approx 2 \times 10^{15}\Gamma \rm \ Hz$ \citep{2008MNRAS.386..945T} and $3\times10^{13}\Gamma \rm \ Hz$ \citep{2007ApJ...660..117C} in the jet comoving frame, respectively. In the EC process, if the energy density of the BLR dominates the EC process is named EC/BLR, otherwise it is named EC/DT. The energy density of BLR ($u_{\rm BLR}$) and dust torus ($u_{\rm DT}$) emission can be approximated \citep{2012ApJ...754..114H} by
    \begin{equation}
    u_{\rm BLR} = \frac{\eta_{\rm BLR}\Gamma^{2}L_{\rm d}}{3\pi r_{\rm BLR}^{2}c[1+\left(r/r_{\rm BLR}\right)^{3}]}, 
    \end{equation}
    and
    \begin{equation}
    u_{\rm DT} = \frac{\eta_{\rm DT}\Gamma^{2}L_{\rm d}}{3\pi r_{\rm DT}^{2}c[1+\left(r/r_{\rm DT}\right)^{4}]}, 
    \end{equation}
    where $r$ is the distance between the central black hole and the dissipation region. $\eta_{\rm BLR} = 0.1$ and $\eta_{\rm DT} = 0.1$ are the the fractions of the disk liminosity $L_{\rm d}$ reprocessed into BLR and dust torus radiation, respectively, $r_{\rm BLR} = 0.1\left(L_{\rm d}/10^{46}\rm erg \ s^{-1}\right)^{1/2} pc$ and $r_{\rm DT} = 2.5\left(L_{\rm d}/10^{46}\rm erg \ s^{-1}\right)^{1/2}\,$pc are the characteristic distances where the above processes taking place. 
    
    Note that these radiation field can also interact with relativistic protons via the Bethe-Heitler pair production and photopion production. We follow the semi-analytical method developed by \citet{2008PhRvD..78c4013K} for the generated spectra of secondary particles in these two processes.  The cooling timescale of both electrons and protons via the aforementioned processes in the three models, which will be discussed in the following section, are shown in Figure 1 for references.
    
    The maximum protons Lorentz factor in the emission region can be approximated as
    \begin{equation}
     \begin{aligned}
    \gamma_{\rm p, max} &= \sqrt{\frac{80}{9}}\frac{eBR}{\alpha m_{ p }c^{2}},  \\
    &\simeq 10^{9} \left(\frac{\alpha}{10}\right)^{-1}\left(\frac{B}{1\rm \ G}\right)\left(\frac{R}{10^{16} \rm cm}\right), \
    \end{aligned}
    \label{f6} 
    \end{equation}
    where $\alpha$ is the parameter which in the case of shock acceleration depends on the spectrum of magnetic turbulence and on the velocity of the upstream-flow \citep{2011IAUS..275...59S}. $\alpha = 10$ for mildly relativistic shocks is adopted in our model \citep{1983A&A...125..249L}. The kinetic luminosity of the magnetic field and nonthermal particles can be approximated as \citep{Celotti_2008}
    \begin{equation}
    L_{k,i} = \pi R^{2}\Gamma^{2}cU_{i}, 
    \end{equation}
    where $U_{i}$ is the energy density of magnetic field or nonthermal particles, here i = $\left\{\rm B, \rm e ,\rm p\right\}$, represents magnetic field, electrons and protons respectively. The best-fit and uncertainty of the model parameters are derived via the Markov Chain Monte Carlo (MCMC) method \citep{emcee_2013}.

    \begin{figure}[ht!]
    \plotone{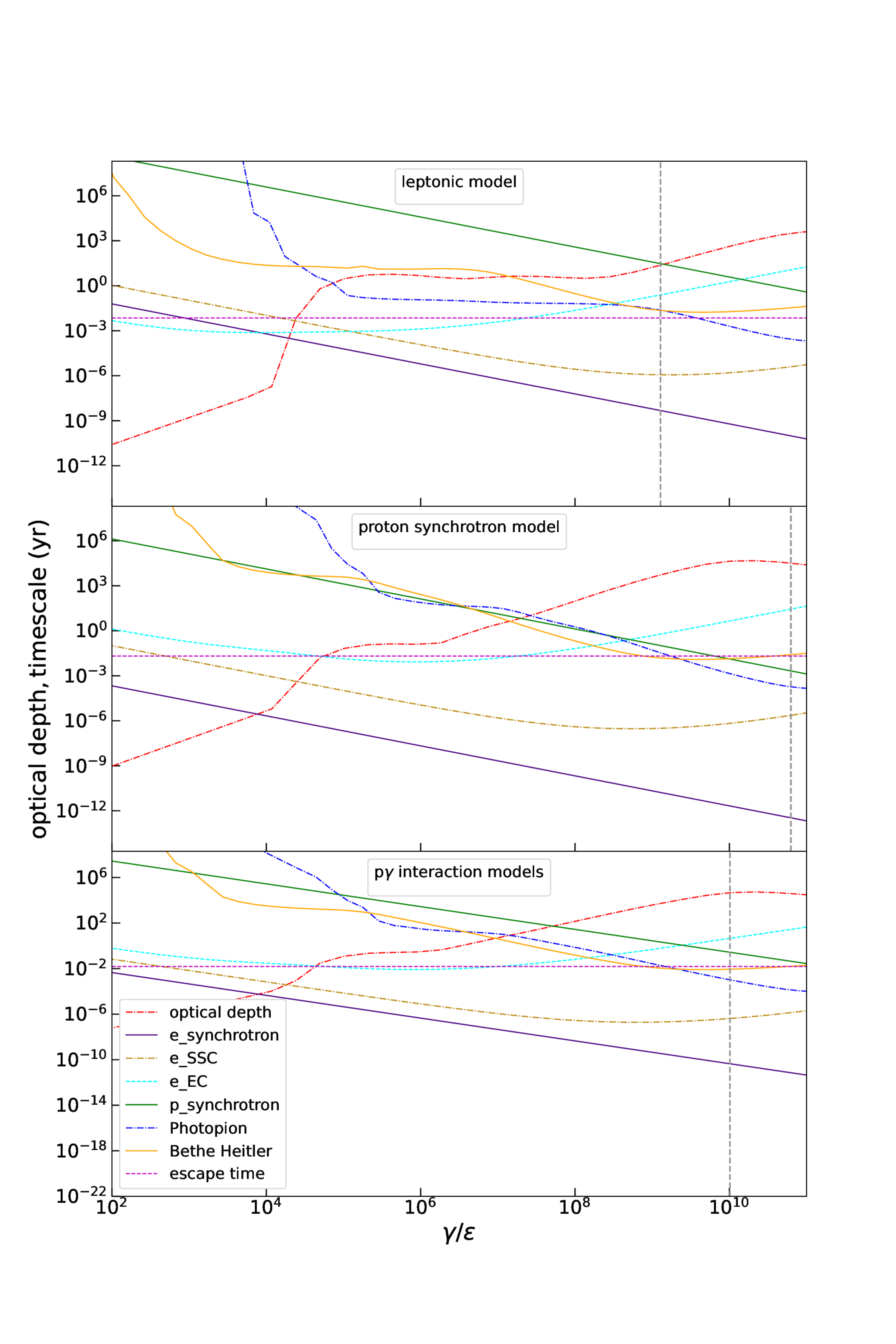}
    \caption{The cooling timescale of protons and electrons under different models, as well as the opacity of the intrinsic $\gamma\gamma$ absorption of the radiation zone. The horizontal axis $\gamma$ represents Lorentz factor of protons for synchrotron, photopion and Bethe-Heitler process in the comoving frames, and also represents Lorentz factor of electrons for synchrotron, SSC and EC process in the comoving frames. In addition, we also plot the optical depth of photons(the red dot-dashed line) in the figure, the $\epsilon = E_{\gamma}/m_{\rm e}c^{2}$, here $E_{\gamma}$ is photon energy. The vertical grey dashed lines show the maximum proton energy allowed by the Hillas condition. Parameters of each model are described in Table 1-3, respectively.
    \label{fig:general}}
    \end{figure}
    
\section{SED Modeling and results}

\subsection{Scenario I: the leptonic model}

    In this scenario, the synchrotron, SSC, EC radiation from an electron population and a simple blackbody emission of disk are used to reproduce the broadband SED of jet. The radiating electrons are assumed to have the number distribution described in Equation (1). The minimum and maximum Lorentz factor of electrons are taken $\gamma_{\rm e,min}$ = 1 and  $\gamma_{\rm e,max} = 1.0\times10^{4}$. The SED from radio to ultraviolet can be attributed by a synchrotron radiation of an electron population and a simple blackbody emission of disk, and the X-ray to $\gamma$-ray can be reproduced by the inverse compton scattering of BLR . 
    \begin{figure}[ht!]
    \plotone{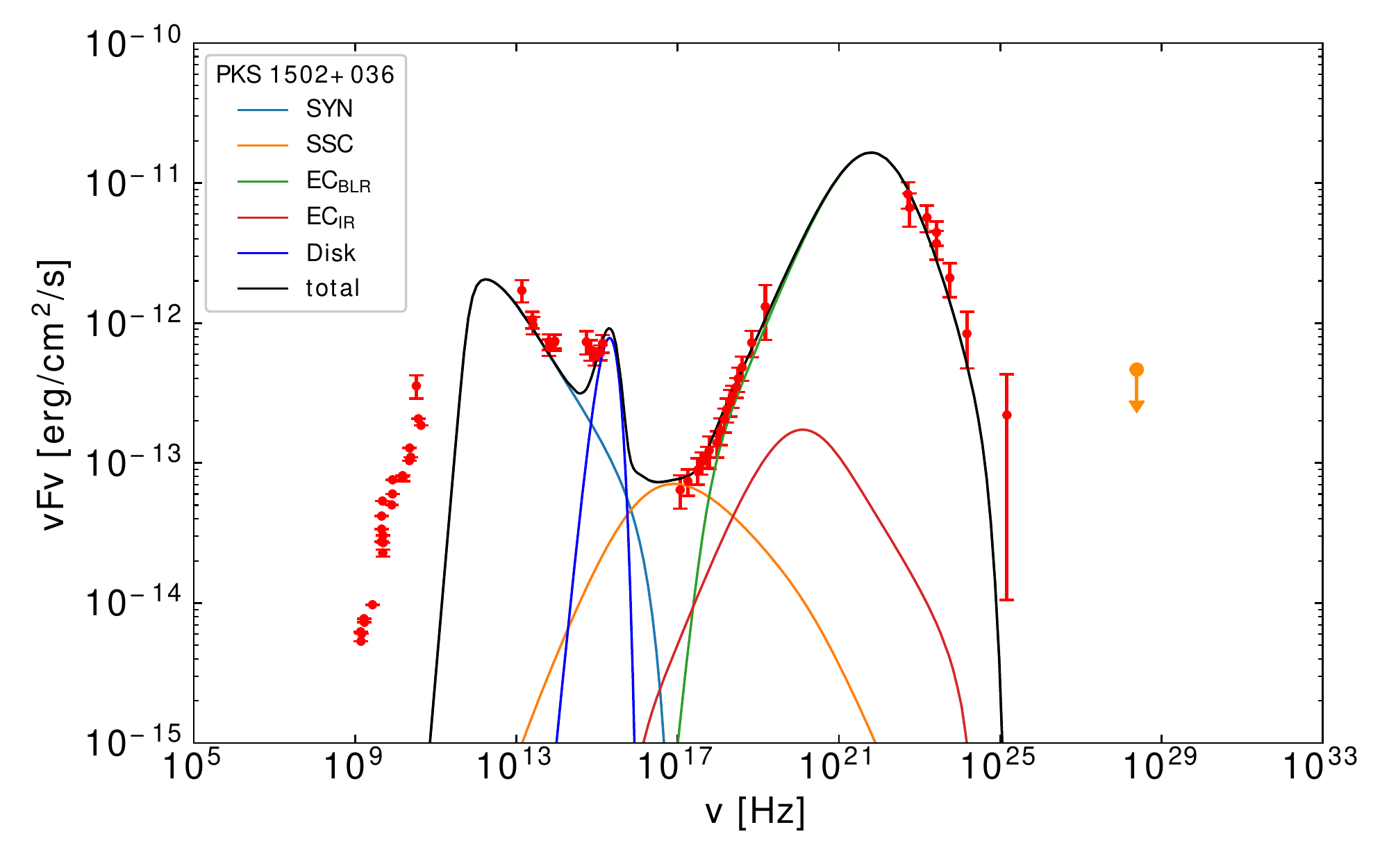}
    \caption{Observed broadband SED of the jet with the leptonic model fits. The Optical-Ultraviolet data and the X-ray to $\gamma$-ray data are taken from \citet{Paliya_2019}, the rest of the data are from NED, the upper limit on neutrino observation is taken from \citet{IceCube_2020}. \label{fig:general}}
    \end{figure}
    The free parameters of the SED modeling are $\Gamma$, $n_{1}$, $n_{2}$, $\gamma_{\rm break}$, $L_{\rm e,inj}$, $r$, $B$. If we assume that the electron energy density $\rm U_e$ is equal to magnetic field $\rm U_B$, then we can get the equipartition magnetic field strength $B_{\rm eq}$ is about 2.2 G. The SED fitting results are shown in Figure 2 and the derived parameters are listed in Table 1. It is found that the single zone leptonic model can reproduce the broadband SED of PKS 1502+036 well, with $\chi^{2}/dof$ being 0.95. 
    \begin{deluxetable*}{ccccccccccc}
    \tablenum{1}\tablecaption{The physical parameters derived from the leptonic model fits\label{tab:messier}}
    \tablewidth{0pt}
    \rule{0pt}{10pt} 
    \tablehead{ 
       \colhead{$\Gamma$} & \colhead{$n_{1}$} & \colhead{$n_{2}$} & \colhead{$\log \gamma_{\rm break}$} & \colhead{$\log L_{\rm e,inj}$} & \colhead{$\log r$} &\colhead{$\log B$} & \colhead{$B_{\rm eq}$} & \colhead{$ L_{k,\rm B} $} & \colhead{$ L_{k,\rm e} $} & \colhead{$\chi^{2}/dof$}\\
       \colhead{} & \colhead{} & \colhead{} & \colhead{} & \colhead{($\rm erg \ \rm s^{-1}$)} & \colhead{(pc)} &\colhead{(G)} & \colhead{(G)} & \colhead{($\rm erg \ \rm s^{-1}$)}&\colhead{($\rm erg \ \rm s^{-1}$)}&\colhead{}
       }
    \decimalcolnumbers
    \startdata
    16.0$_{-3.7}^{+3.8}$ & 1.4$_{-0.5}^{+0.3}$ & 3.8$_{-0.3}^{+0.4}$ & 2.1$_{-0.3}^{+0.3}$ & 42.2$_{-0.3}^{+0.6}$ & -1.4$_{-0.3}^{+0.3}$  & 0.3$_{-0.2}^{+0.2}$ & 2.2 & 1.5e44 & 1.9e44 & 0.95\\
    \enddata
    \tablecomments{The column information are as follows: Col.(1) the bulk Lorentz factor; Col.(2) low energy spectral index of electron; Col.(3) high energy spectral index of electron; Col.(4) The break Lorentz factor; Col.(5) the injection luminosity of electrons;  Col.(6)  the distance between black hole and blob; col.(7) The magnetic field strength in SED fitting; col.(8) The equipartition magnetic field strength; col.(9) the luminosity of magnetic field; col.(10) the luminosity of electrons; col.(11) the reduced $\chi^{2}$, the $dof$ is degree of freedom. The luminosity of disk is fixed to $6.03\times10^{44}\rm erg \ \rm s^{-1}$\citep{Paliya_2019}.}
    \end{deluxetable*} 

\subsection{Scenario II: proton synchrotron model}

    It was suggested that the X-ray emission and even $\gamma$-ray emission of jets in some blarzars may be explained by the synchrotron radiation of protons \citep{Aharonian_2000,2014_Kundu,2013_Boettcher}. On the other hand, \citet{2013_Boettcher} selected 12 Fermi-LAT-detected blazars and modeled their broadband SEDs, but found the proton synchrotron model has difficulty describing the GeV break in the SEDs of two FSRQs while provides appropriate fits for all other blazars in their sample. This model requires very large powers in relativistic protons, i.e., $L_{\rm p} \sim 10^{47} - 10^{49} \rm \ erg\ s^{-1}$. This value is close to or even exceeds the Eddington luminosity, in most cases dominating the total power in the jet \citep{2013_Boettcher}, and this model needs a larger magnetic field strength generally. \citet{2013_Boettcher} found that the magnetic field strength is generally larger than $10\ \rm G$ in order to explain X-ray to $\gamma$-ray emission via synchrotron radiation of protons \citep{2013_Boettcher}. Whether such a strong magnetic field can be achieved in AGN jet remains unclear and hence the proton synchrotron model is not confirmed yet. 

    In the proton synchrotron model, the low-energy emission still comes from electrons, which is the same as the leptonic model. The minimum and maximum Lorentz factor of electrons are set to $\gamma_{\rm e,min}$ = 1, $\gamma_{\rm e,max}$ = $1.0\times10^{4}$, and the minimum and maximum Lorentz factor of protons are $\gamma_{\rm p,min}$ = 1, and $\gamma_{\rm p,max} = 6.3\times10^{10}$, respectively. The maximum Lorentz factor of protons are approximated by Equation (6). By fitting the optically thin spectrum, \citet{D_Ammando_2013} found that the rest-frame brightness temperature is $T^{\prime}_{\rm B}\sim2.5\times10^{13}~\rm K$, which exceeds the value derived for the Compton catastrophe. Assuming that such a high value is due to Doppler boosting, they estimated the variability Doppler factor $\delta$ = 6.6. This value is smaller than the Doppler factor obtained in \citet{2009ApJ...707L.142A_Abdo} by modelling the SED ($\delta$ = 18). To make the IC radiation of electrons negligible, we here adopt this small Doppler factor ($\delta$ = 6.6) to suppress the external radiation energy density in the jet's comoving frame. 
    

    The SED fitting results are shown in Figure 3, and the fitting parameters are shown in Table 2. The corresponding $\chi^{2}/dof$ of the fitting is 1.84, which is larger than the $\chi^{2}/dof$ under the leptonic model. Meanwhile we find the break Lorentz factor of proton is $\sim 1.3 \times 10^{8}$, this value means that protons in the emission region are accelerated efficiently. In conclusion, proton synchrotron model is still controversial, new observational evidence would be needed to support or rule out this model. The free parameters of the SED modeling in this model are $B$, $\gamma_{\rm break}$, $n_{1}$, $n_{2}$, $L_{\rm e,inj}$, $L_{\rm p,inj}$, $r$, $\alpha$, $\beta$, $\gamma_{\rm p,break}$. 
    
    \begin{figure}[ht!]
    \plotone{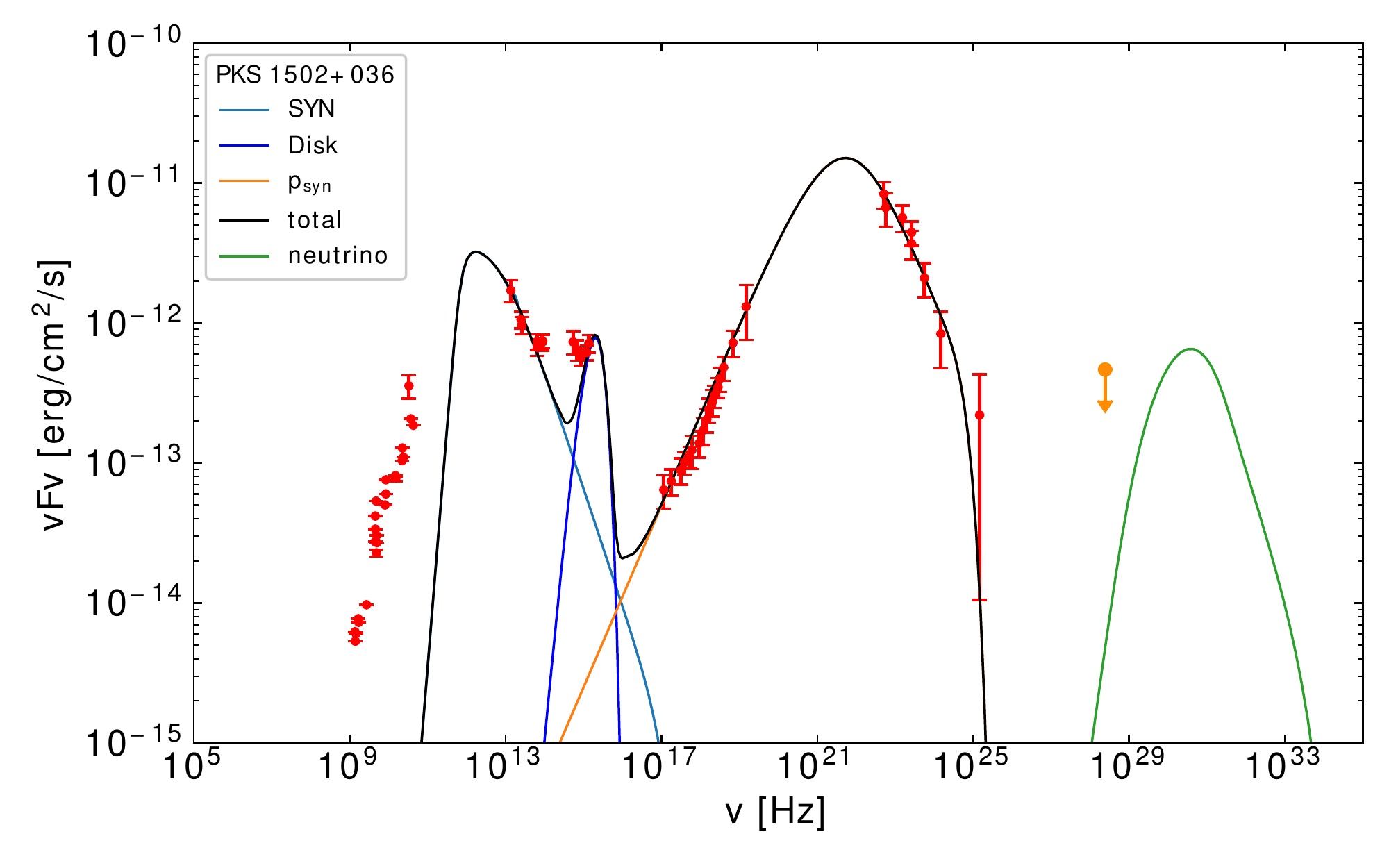}
    \caption{Observed broadband SED of the jet with the proton synchrotron model fits. \label{fig:general}}
    \end{figure}
    
    \begin{deluxetable*}{cccccccccccccc}
    \tablenum{2}
    \tablecaption{The physical parameters derived from the proton synchrotron model fits \label{tab:messier}}
    \tablewidth{0pt}
    \rule{0pt}{10pt}
    \tablehead{
        \colhead{$n_{1}$} & \colhead{$n_{2}$} & \colhead{$\log \gamma_{\rm break}$} & \colhead{$ \log L_{\rm e,inj}$} & \colhead{$\log r$} & \colhead{$\log B$}& \colhead{$\log L_{\rm p,inj}$} & \colhead{$\alpha$} & \colhead{$\beta$}& \colhead{$\log \gamma_{\rm p,break}$} & \colhead{$ B_{\rm eq}$} & \colhead{$L_{k,\rm B}$} & \colhead{$L_{k,\rm p+e}$} & \colhead{$\chi^{2}/dof$}  \\
        \colhead{} & \colhead{} & \colhead{} & \colhead{($\rm erg \ \rm s^{-1}$)} & \colhead{(pc)} & \colhead{(G)} & \colhead{($\rm erg \ \rm s^{-1}$)} & \colhead{} &\colhead{} & \colhead{} & \colhead{(G)} & \colhead{($\rm erg \ \rm s^{-1}$)}&\colhead{($\rm erg \ \rm s^{-1}$)}&\colhead{}
        }
    \decimalcolnumbers
    \startdata
    1.5$_{-0.24}^{+0.2}$ & 3.7$_{-0.3}^{+0.3}$ & 1.9$_{-0.1}^{+0.1}$ & 42.9$_{-0.1}^{+0.1}$ & -0.9$_{-0.1}^{+0.1}$ & 1.5$_{-0.2}^{+0.1}$  & 45.6$_{-0.2}^{+0.2}$ & 1.7$_{-0.1}^{+0.1}$ & 4.2$_{-0.2}^{+0.2}$ & 8.1$_{-0.1}^{+0.1}$ & 48.0 & 2.0e46 & 4.0e46 & 1.84\\
    \enddata
    \tablecomments{The column information are as follows: (1) low energy spectral index of electrons; Col.(2) high energy spectral index of electrons; Col.(3) The break Lorentz factor of electrons; Col.(4) the injection luminosity of electrons;  Col.(5) the distance between black hole and blob; Col.(6) the magnetic field in SED fitting; Col.(7) the injection luminosity of protons; Col.(8) low energy spectral index of protons; Col.(9) high energy spectral index of protons; Col.(10) The break Lorentz factor of protons; Col.(11) the equipartition magnetic field strength; Col.(12) the luminosity of magnetic field; col.(13) the total luminosity of electrons and protons; col.(14) the reduced $\chi^{2}$, the $dof$ is degree of freedom. The luminosity of disk is fixed to $6.03\times10^{44}\rm erg \ \rm s^{-1}$\citep{Paliya_2019}.}
    \end{deluxetable*}
    
    \begin{deluxetable*}{cccccccccccccc}
    \tablenum{3}
    \tablecaption{The physical parameters derived from the $p\gamma$ interaction models fits \label{tab:messier}}
    \tablewidth{0pt}
    \tablehead{
        \colhead{$n_{1}$} & \colhead{$n_{2}$} & \colhead{$\log \gamma_{break}$} & \colhead{$\log L_{\rm e,inj}$} & \colhead{$\log r$} & \colhead{$\log B$}& \colhead{$\log L_{\rm p,inj}$} & \colhead{$\alpha$} & \colhead{$\beta$}& \colhead{$\log \gamma_{p,break}$} & \colhead{$B_{\rm eq}$} & \colhead{$L_{k,\rm B}$} & \colhead{$L_{k,\rm p+e}$} & \colhead{$\chi^{2}/dof$}\\
        \colhead{} & \colhead{} & \colhead{} & \colhead{($\rm erg \ \rm s^{-1}$)} & \colhead{(pc)} & \colhead{(G)} & \colhead{($\rm erg \ \rm s^{-1}$)} & \colhead{} &\colhead{} & \colhead{} & \colhead{(G)} & \colhead{($\rm erg \ \rm s^{-1}$)}&\colhead{($\rm erg \ \rm s^{-1}$)}&\colhead{}
        }
    \decimalcolnumbers
    \startdata
    1.5$_{-0.2}^{+0.2}$ & 3.1$_{-0.2}^{+0.2}$ & 2.0$_{-0.1}^{+0.2}$ & 43.1$_{-0.1}^{+0.1}$ & -1.0$_{-0.1}^{+0.1}$ & 0.9$_{-0.1}^{+0.1}$  & 46.3$_{-0.2}^{+0.1}$ & 1.3$_{-0.1}^{+0.1}$ & 4.2$_{-0.2}^{+0.2}$ & 7.5$_{-0.1}^{+0.1}$ & 147.1 & 5.2e44 & 2.1e47 & 2.96 \\
    \enddata
    \tablecomments{The column information are as follows: (1) low energy spectral index of electrons; Col.(2) high energy spectral index of electrons; Col.(3) The break Lorentz factor of electrons; Col.(4) the injection luminosity of electrons;  Col.(5) the distance between black hole and blob; Col.(6) the magnetic field in SED fitting; Col.(7) the injection luminosity of protons; Col.(8) low energy spectral index of protons; Col.(9) high energy spectral index of protons; Col.(10) The break Lorentz factor of protons; Col.(11) the equipartition magnetic field strength; Col.(12) the luminosity of magnetic field; col.(13) the total luminosity of electrons and protons; col.(14) the reduced $\chi^{2}$, the $dof$ is degree of freedom. The luminosity of disk is fixed to $6.03\times10^{44}\rm erg \ \rm s^{-1}$\citep{Paliya_2019}.}
    \end{deluxetable*}

\subsection{Scenario III: Photopion and Bethe-Heitler model}

    Interactions between high-energy protons and the radiation field of the source have been widely considered as radiation channels of protons in AGN. One of the main processes is the photopion process, i.e.,
    \begin{equation}
    \begin{aligned}
    p& + \gamma \rightarrow p^{\prime} + \pi^{0}, \\
    p& + \gamma \rightarrow n + \pi^{+}, \\
    p& + \gamma \rightarrow p^{\prime} + \pi^{+} + \pi^{-}, \
    \end{aligned}
    \end{equation}
    via which unstable pions are produced and further decay into
    \begin{equation}
    \begin{aligned}
    \pi^{0}& \rightarrow 2\gamma, \\
    \pi^{+} \rightarrow \mu^{+} + \nu_{\mu}& \rightarrow e^{+} + \nu_{e} + \Bar{\nu_{\mu}} + \nu_{\mu},\\
    \pi^{-} \rightarrow \mu^{-} + \Bar{\nu_{\mu}}& \rightarrow e^{-} + \bar\nu_{e} + \nu_{\mu} + \Bar{\nu_{\mu}},\
    \end{aligned}
    \end{equation}
    Another important process is the Bethe–Heitler pair-production, leading to the production of electron/positron pairs, i.e.,
    \begin{equation}
    \begin{aligned}
    p + \gamma \rightarrow p^{\prime} + e^{+} + e^{-}, \
    \end{aligned}
    \end{equation}
    
    In these processes, radiation fields of the AGN serve as the targets, including blackbody emission of the BLR and the dusty torus, the synchrotron radiation of primary electrons, as well as the radiation of secondary pairs developed in the electromagnetic cascade initiated by the two processes. The photopion process and the Bethe-Heitler process including the cascade are calculated following the method shown in \citet{Wang_2022}.
    
    
    We can find out the best-fit parameters in this model with the MCMC method, and get the best fit to the SED (see Appendix for the associated MCMC results). As shown in Figure 4, however, the resulting spectral shape cannot coincide with the data at all. This is due to the nature of the cascade emission. At the MeV-GeV band, the emission is dominated by pairs generated in the cascade process. The spectrum is somewhat independent on the initial parameters as long as the cascade is sufficiently developed.
    
    The minimum and maximum Lorentz factor of electrons are $\gamma_{\rm e,min}$ = 1, $\gamma_{\rm e,max}$ = $1.0\times10^{4}$, and the minimum and maximum Lorentz factor of protons are $\gamma_{\rm p,min}$ = 1, and $\gamma_{\rm p,max} = 1.0\times10^{10}$ respectively. Here $B$ is magnetic field strength. The maximum Lorentz factor of protons are approximated by Equation (6). In $p\gamma$ interaction models, we adopt $\delta$ = 6.6 \citep[the variability Doppler factor in ][]{D_Ammando_2013}. The free parameters of the SED modeling are $B$, $\gamma_{\rm break}$, $n_{1}$, $n_{2}$, $L_{\rm e,inj}$, $L_{\rm p,inj}$, $r$, $\alpha$, $\beta$, $\gamma_{\rm p,break}$. The SED fitting results are shown in Figure 4. The fitting parameters are listed in Table 3. The derived $\chi^{2}/dof$ is 2.96. The result means that the $p\gamma$ interaction models cannot reproduce the high energy band SED of PKS 1502+036. 

    \begin{figure}[ht!]
    \plotone{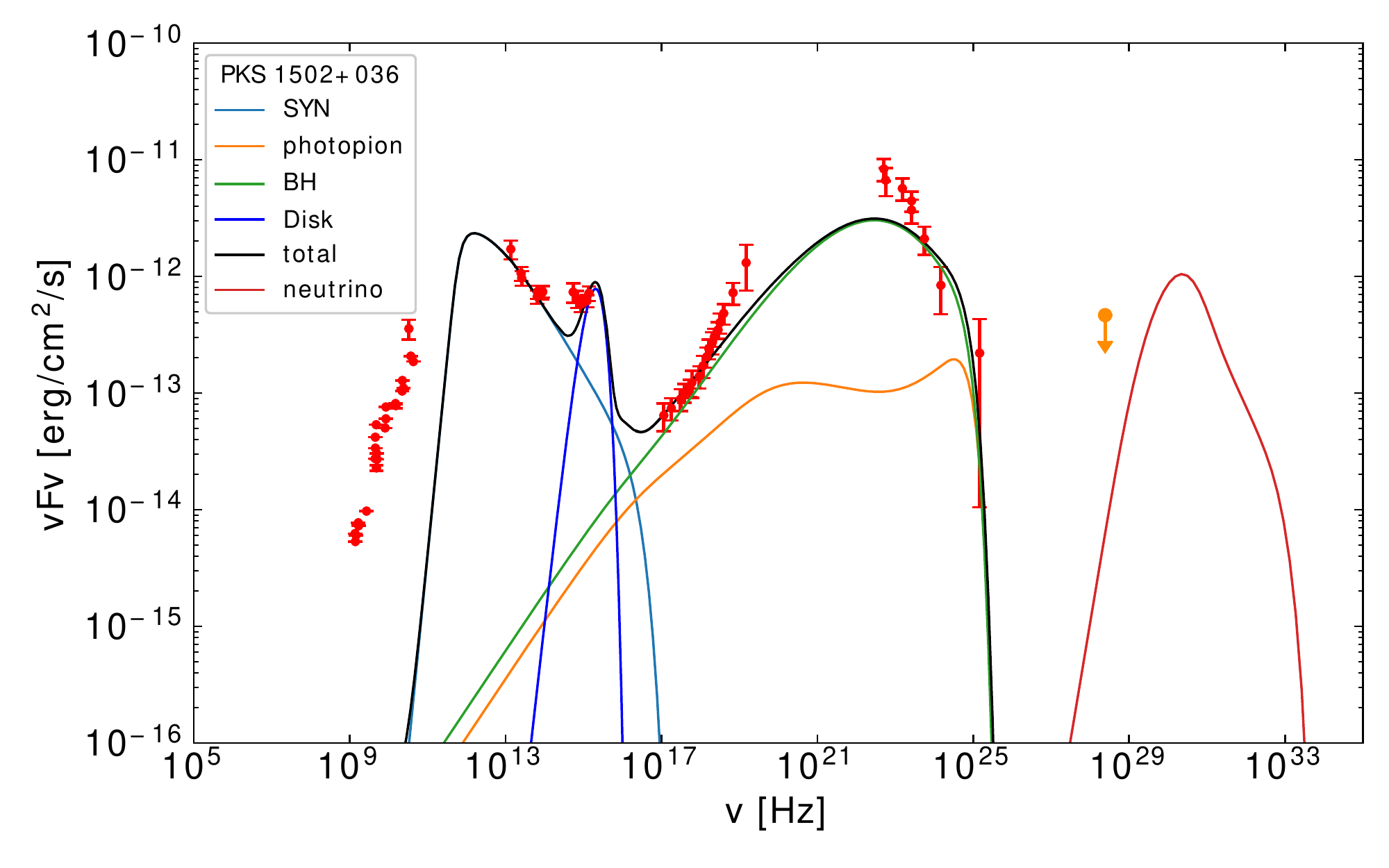}
    \caption{Observed broadband SED of the jet with $p\gamma$ interactions including the emission of pairs generated in the electromagnetic cascade initiated by these processes. \label{fig:general}}
    \end{figure}

\section{The maximum injection luminosity of proton under the leptonic model}

    As described above, we have attempted to use leptonic model and lepto-hadronic model to explain the broad SED of jet in PKS~1502+036. We find the leptonic model is a popular model to explain the broad SED of jet, and the $p\gamma$ interaction model cannot explain the high energy SED of jet. Though the photopion and Bethe-Heitler model cannot explain the high energy SED of jet, high energy protons may still exist in the jet of PKS~1502+036. So we attempt to use the broadband SED of jet to limit the maximum injection luminosity of proton under the leptonic model.

    In the leptonic model case, we can assume a power-law distribution of protons, with the adopted spectrum index of protons to be 1.5, 2.0 and 2.5, respectively. The corresponding results of SED fitting are shown in Figure~5. In this work, we use p-value to limit the maximum injection luminosity of protons. A smaller p-value indicates that the probability of the event happening is lower. In detail, we define the injected luminosity of proton as the maximum luminosity when the cumulative probability of the $\chi^2$ distribution reaches 3$\sigma$ (0.997), the corresponding $p$ value is 0.003. Given the degrees of freedom in our model is 21,  the corresponding $\chi^{2}/dof$ is 2.07. Then we gradually increase the injection luminosity of protons under the leptonic model and the $\chi^{2}/dof$ will change gradually. We can get the maximum injection luminosity of proton when the $\chi^{2}/dof$ is increased by 2.07. 
    
   The minimum proton Lorentz factor is set to 1 and the maximum proton Lorentz factor is $1.3\times10^{9}$, which is approximated by Equation (6). Table~4 shows that the maximally allowed proton injection luminosity increases from $10^{43.0}$ $\rm erg/s$ to $10^{45.6}$ $\rm erg/s$ when the spectral index increases from 1.5 to 2.5.
   
   It is worth noting that the main constraint on the proton luminosity comes from the soft X-ray flux. This is because the synchrotron radiation of electron/positron pairs generated in the EM cascade peak around the soft X-ray band, while it is also the ``valley'' in the SED at the soft X-ray band. This is the reason why the main deviation between theoretical flux and observed flux is at the soft X-ray band as shown in Figure~5.

   \begin{deluxetable*}{cccccc}
   \tablenum{4}
   \tablecaption{The maximum injection luminosity of protons under the leptonic model\label{tab:messier}}
   \tablewidth{0pt}
   \tablehead{
      \colhead{the spectrum index of protons}&\colhead{$L_{\rm p,inj}$} &\colhead{$\chi^{2}/dof$} &\colhead{$L_{k,\rm B}$} &\colhead{$L_{k,\rm p+e}$} & \colhead{$L_{k,\rm p(\textgreater1~EeV)}$} \\
     \colhead{} & \colhead{($\rm erg \ \rm s^{-1}$)} & \colhead{} & \colhead{($\rm erg \ \rm s^{-1}$)} & \colhead{($\rm erg \ \rm s^{-1}$)} & \colhead{($\rm erg \ \rm s^{-1}$)}
     }
   \decimalcolnumbers
   \startdata
   1.5 & $10^{43.0}$ & 3.02 & 1.5e44 & 1.8e45 & 1.2e44\\
   2.0 & $10^{43.3}$ & 3.02 & 1.5e44 & 3.4e45 & 2.5e43\\
   2.5 & $10^{45.6}$ & 3.02 & 1.5e44 & 7.5e47 & 1.8e42\\
   \enddata
   \tablecomments{The column information are as follows: (1) the spectral index of protons; Col.(2) the maximum injection luminosity of protons in the comoving frames; Col.(3) the reduced $\chi^{2}$, the $dof$ is degree of freedom; col.(4) the luminosity of magnetic field; col.(5) the total luminosity of electrons and protons; col.(6) the luminosity of protons ($\textgreater1~\rm EeV$)}
  \end{deluxetable*}

\begin{figure}[ht!]
\plotone{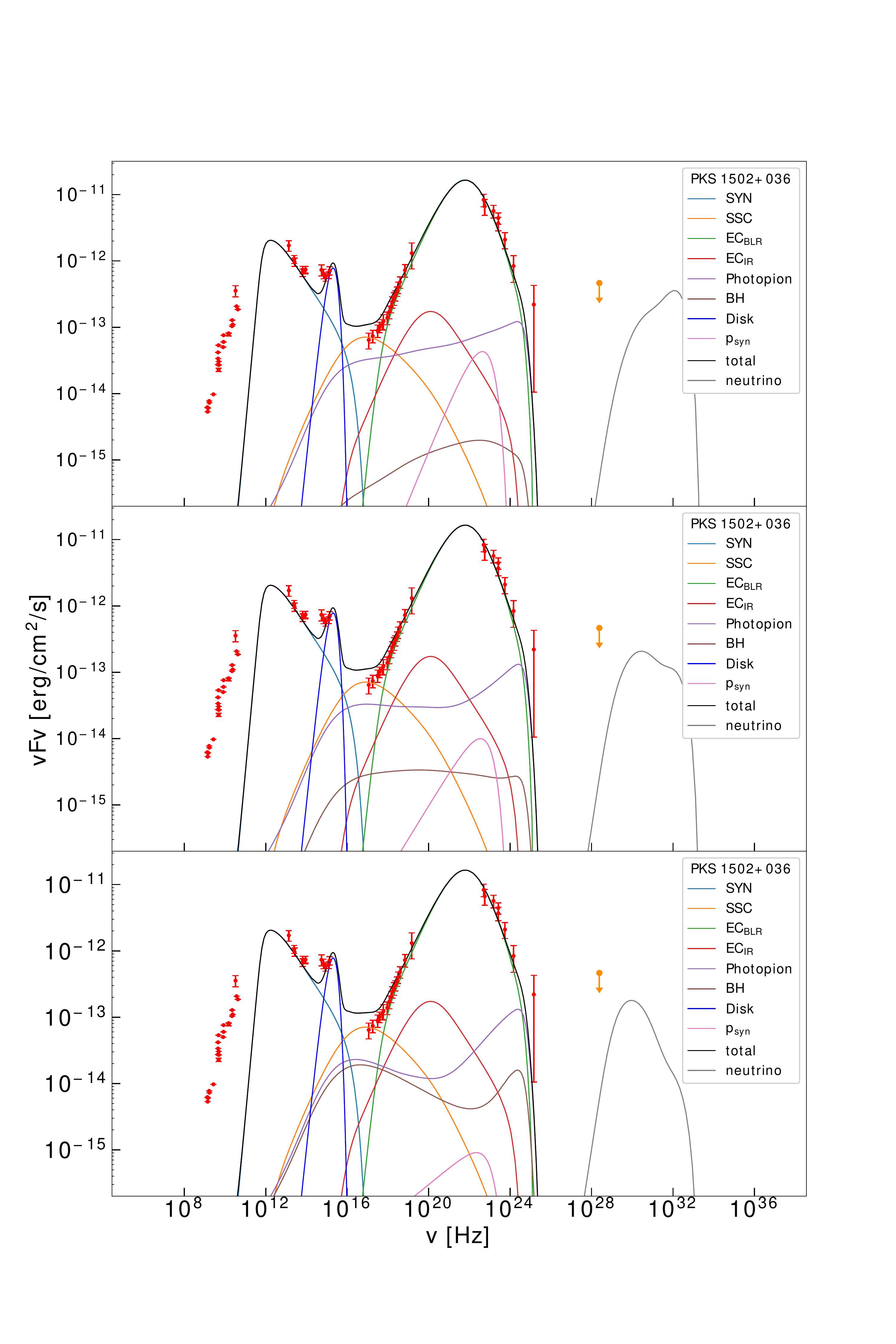}
\caption{Observed broadband SED of the jet with the leptonic model fits. The spectrum index of protons is 1.5, 2.0 and 2.5 respectively from the top to the bottom of the picture. \label{fig:general}}
\end{figure}

\section{Discussion and Summary}

   In this work, we study the gamma-ray radiation mechanism of RL-NLS1 PKS~1502+036 in the framework of both the leptonic model and the lepto-hadronic model. In both models, the low-energy radiation are ascribed to the synchrotron radiation of electrons, but the former employs the inverse Compton radiation of electrons for the X-ray to $\gamma$-ray emission while the latter attributes the high-energy emission to protons. The two scenarios are considered in the lepto-hadronic models, namely, the proton synchrotron model and the proton-photon interaction model (including the photopion production and the Bethe-Heitler process) respectively. Our calculation showed that the leptonic model can reproduce the broadband SED of this source best with the smallest value for $\chi^2/dof$ (0.95) among three models, implying that the inverse Compton radiation is the favorable mechanism for the production of the gamma-ray emission of PKS~1502+036. 
   The same conclusions have been drawn in previous works on
   the same object \citep{Paliya_2016,Paliya_2019}, in which the leptonic radiative processes have been favored. In the leptonic model, the derived Doppler factor $\delta$ = 18.6 in our work. This value is comparable to the one reported in \citet{Paliya_2019}, which is about 17.2. The derived distance between the black hole and the blob is about 0.04 pc in our work, this value is lower than the value in \citet{Paliya_2016} and \citet{Paliya_2019}, which is 0.16 pc and 0.126 pc, respectively. The derived magnetic field strength in our work is about 2 G, larger than the value in \citet{Paliya_2019}, which is about 0.25 G.
   
    The proton synchrotron model could also yield an acceptable fitting to the high energy emission with $\chi^2/dof=1.84$. However, the required kinetic proton luminosity exceeds the Eddington luminosity of this object by about one order of magnitude, which is $L_{\rm Edd} \approx 4\times 10^{45}\left(M_{\rm BH}/10^{7.6} M_{\odot} \right) \rm erg\ s^{-1}$, with $M_{\rm BH}$ $\approx$ $10^{7.6}$ $M_{\odot}$ being the supermassive black hole mass of this object \citep{Paliya_2019}. Given the low efficiency of synchrotron radiation of protons, the required magnetic field strength is around 34~G. It is quite large for the general value of parsec-scale AGN jet, although comparable with the equipartition magnetic ﬁeld strength which is about 48~G in this case. As a consequence, the derived kinetic luminosity of the magnetic field is $2\times10^{46} \rm erg \ s^{-1}$, also exceeding the Eddington luminosity by a factor of 5. This makes the proton synchrotron model dubious because whether such an extraordinary accretion rate can be stable for a long time is unclear. On the other hand, the proton-photon interaction model fails to explain the X-ray to $\gamma$-ray emission, because the spectral shape of the electromagnetic cascade initiated by these interactions is much flatter than the observed data. Besides, the required proton power also significantly exceeds the Eddington luminosity in order to make the generated X-ray/$\gamma$-ray flux at the comparable level of the observed one. 
    
    It may be worth mentioning that, although the lepto-hadronic model is not favored, it does not mean protons are not accelerated in the jet of NLS1 galaxy. In fact, it is widely believed that ultra-high-energy cosmic rays (UHECRs) above the so-called ``ankle'' (i.e, at a few EeV), where the composition is mainly proton \citep{Abbasi_2017, Schroder_2019}, predominantly originate from extra-galactic sources \citep{Pierre_2013, Aab_2018}, such as AGNs. We then estimate the largest amount of protons that can be allowed to be injected in the jet under the leptonic model. To do this, we added a proton component, assuming a power-law distribution with the spectral index being 1.5, 2.5 and 2.5 respectively, in the leptonic model and calculate the overall flux given by both electrons and protons (mainly from the induced EM cascades). The maximum injection luminosity of protons can be obtained when the overall flux violates the observation at $3\sigma$ confidence level (i.e., when $\chi^{2}/dof$ is increased by 2.07 in our case). The obtained largest kinetic luminosity of protons are $L_{p,k}=1.8\times 10^{45}\rm erg~s^{-1}$, $3.4\times 10^{45}\rm erg~s^{-1}$, and $7.5\times 10^{47}\rm erg~s^{-1}$ for $n_p=1.5, ~2.0, ~2.5$, respectively.
   
    Note that NLS1 galaxies constitute a fraction of $\sim 15\%$ of Seyfert 1 galaxies \citep{2008RMxAC..32...86K}, whereas the spatial density of the latter at the local universe is about $3\times 10^{-4}\rm Mpc^{-3}$ \citep{1992ApJ...393...90H}. Therefore, if we may estimate the upper limit of the local UHECR energy production of NLS1 galaxies if using the result for PKS~1502+036 as the
    representative, yielding $\dot{W}_p\simeq 1.7\times 10^{47}\rm erg~Mpc^{-3}yr^{-1}$ for $n_p=1.5$, $\dot{W}_p\simeq 3.5\times 10^{46}\rm erg~Mpc^{-3}yr^{-1}$ for $n_p=2.0$, and $\dot{W}_p\simeq 2.6\times 10^{45}\rm erg~Mpc^{-3}yr^{-1}$ for $n_p=2.5$. The constrained upper limits of the energy production rate are larger than the required value to explain the measured UHECRs spectrum at ankle, i.e., $\sim 10^{44}\rm erg~Mpc^{-3}yr^{-1}$ \citep[e.g.,][]{2009JCAP...03..020K, 2019FrASS...6...23B}. Therefore, NLS1 galaxies in principle may contribute the measured UHECRs flux at the ankle. However, even so, the maximally achievable proton energy is just above a few EeV (see Figure 1) and hence their contribution cannot continue to higher energies, according to the best-fit model parameters (i.e., the magnetic field and the size of the radiation zone) and the Hillas condition \citep{1984ARA&A..22..425H}.

\begin{acknowledgments}

We sincerely thank the anonymous referee for the constructive comments and helpful suggestions. ZJW and JFW acknowledge support by the National Science Foundation of China (NSFC) grants U1831205, 12033004 and 12221003. RYL acknowledges NSFC grant U2031105. This research has made use of the NASA/IPAC Extragalactic Database, which is funded by the National Aeronautics and Space Administration and operated by the California Institute of Technology.

\end{acknowledgments}

\vspace{5mm}



\bibliography{sample631}

\begin{thebibliography}{}
\expandafter\ifx\csname natexlab\endcsname\relax\def\natexlab#1{#1}\fi
\providecommand{\url}[1]{\href{#1}{#1}}
\providecommand{\dodoi}[1]{doi:~\href{http://doi.org/#1}{\nolinkurl{#1}}}
\providecommand{\doeprint}[1]{\href{http://ascl.net/#1}{\nolinkurl{http://ascl.net/#1}}}
\providecommand{\doarXiv}[1]{\href{https://arxiv.org/abs/#1}{\nolinkurl{https://arxiv.org/abs/#1}}}

\bibitem[{Aab {et~al.}(2018)Aab, Abreu, Aglietta, Albuquerque, Albury,
  Allekotte, Almela, Castillo, Alvarez-Mu{\~n}iz, Anastasi, Anchordoqui,
  Andrada, Andringa, Aramo, Asorey, Assis, Avila, Badescu, Balaceanu, Barbato,
  Luz, Baur, Becker, Bellido, Berat, Bertaina, Bertou, Biermann, Biteau,
  Blaess, Blanco, Blazek, Bleve, Boh{\'a}{\v c}ov{\'a}, Bonifazi, Borodai,
  Botti, Brack, Bretz, Bridgeman, Briechle, Buchholz, Bueno, Buitink, Buscemi,
  Caballero-Mora, Caccianiga, Calcagni, Cancio, Canfora, Carceller, Caruso,
  Castellina, Catalani, Cataldi, Cazon, Chinellato, Chudoba, Chytka, Clay,
  Cerutti, Colalillo, Coleman, Coluccia, Concei{\c c}{\~a}o, Consolati,
  Contreras, Cooper, Coutu, Covault, Daniel, Dasso, Daumiller, Dawson, Day,
  de~Almeida, de~Jong, Mauro, de~Neto, Mitri, de~Oliveira, de~Souza, Debatin,
  Deligny, Dhital, Castro, Diogo, Dobrigkeit, D'Olivo, Dorosti, dos Anjos,
  Dova, Dundovic, Ebr, Engel, Erdmann, Escobar, Etchegoyen, Falcke, Farmer,
  Farrar, Fauth, Fazzini, Feldbusch, Fenu, Ferreyro, Figueira, Filip{\v c}i{\v
  c}, Freire, Fujii, Fuster, Garc\'{\i}a, Gemmeke, Gherghel-Lascu, Ghia,
  Giaccari, Giammarchi, Giller, G{\l}as, Glombitza, Golup, Berisso, Vitale,
  Gonz{\'a}lez, Goos, G{\'o}ra, Gorgi, Gottowik, Grubb, Guarino, Guedes, Guido,
  Halliday, Hampel, Hansen, Harari, Harrison, Harvey, Haungs, Hebbeker, Heck,
  Heimann, Hill, Hojvat, Holt, Homola, H{\"o}randel, Horvath, Hrabovsk{\'y},
  Huege, Hulsman, Insolia, Isar, Jandt, Johnsen, Josebachuili, Jurysek,
  K{\"a}{\"a}p{\"a}, Kampert, Keilhauer, Kemmerich, Kemp, Klages, Kleifges,
  Kleinfeller, Krause, Kuempel, Mezek, Awad, Lago, LaHurd, Lang, Legumina,
  de~Oliveira, Lenok, Letessier-Selvon, Lhenry-Yvon, Presti, Lopes, L{\'o}pez,
  Casado, Lorek, Luce, Lucero, Malacari, Mallamaci, Mancarella, Mandat,
  Mantsch, Mariazzi, Mari{\c s}, Marsella, Martello, Martinez, Bravo, Mathes,
  Mathys, Matthews, Matthiae, Mayotte, Mazur, Medina-Tanco, Melo, Menshikov,
  Merenda, Michal, Micheletti, Middendorf, Miramonti, Mitrica, Mockler,
  Mollerach, Montanet, Morello, Morlino, Mostaf{\'a}, M{\"u}ller, Muller,
  M{\"u}ller, Mussa, Nellen, Nguyen, Niculescu-Oglinzanu, Niechciol, Nitz,
  Nosek, Novotny, No{\v z}ka, Nucita, N{\'u}{\~n}ez, Olinto, Palatka, Pallotta,
  Papenbreer, Parente, Parra, Pech, Pedreira, Pȩkala, Pelayo,
  Pe{\~n}a-Rodriguez, Pereira, Perlin, Perrone, Peters, Petrera, Phuntsok,
  Pierog, Pimenta, Pirronello, Platino, Poh, Pont, Porowski, Prado, Privitera,
  Prouza, Puyleart, Querchfeld, Quinn, Ramos-Pollan, Rautenberg, Ravignani,
  Reininghaus, Ridky, Riehn, Risse, Ristori, Rizi, de~Carvalho, Rojo,
  Roncoroni, Roth, Roulet, Rovero, Ruehl, Saffi, Saftoiu, Salamida, Salazar,
  Saleh, Salina, S{\'a}nchez, Santos, Santos, Sarazin, Sarmento,
  Sarmiento-Cano, Sato, Savina, Schauer, Scherini, Schieler, Schimassek,
  Schimp, Schmidt, Scholten, Schov{\'a}nek, Schr{\"o}der, Schr{\"o}der,
  Schumacher, Sciutto, Shellard, Sigl, Silli, Sima, {\v S}m\'{\i}da, Snow,
  Sommers, Soriano, Souchard, Squartini, Stanca, Stani{\v c}, Stasielak,
  Stassi, Stolpovskiy, Streich, Suarez, Su{\'a}rez-Dur{\'a}n, Sudholz,
  Suomij{\"a}rvi, Supanitsky, {\v S}up\'{\i}k, Szadkowski, Taboada, Taborda,
  Tapia, Timmermans, Peixoto, Tom{\'e}, Elipe, Travnicek, Trini, Tueros,
  Ulrich, Unger, Urban, Galicia, Vali{\~n}o, Valore, van Bodegom, van~den Berg,
  van Vliet, Varela, C{\'a}rdenas, V{\'a}zquez, Veberi{\v c}, Ventura, Quispe,
  Verzi, Vicha, Villase{\~n}or, Vorobiov, Wahlberg, Wainberg, Watson, Weber,
  Weindl, Wiede{\'n}ski, Wiencke, Wilczy{\'n}ski, Wirtz, Wittkowski,
  Wundheiler, Yang, Yushkov, Zas, Zavrtanik, Zavrtanik, Zehrer, Zepeda,
  Zimmermann, Ziolkowski, Zong, \& Zuccarello}]{Aab_2018}
Aab, A., Abreu, P., Aglietta, M., {et~al.} 2018, The Astrophysical Journal,
  868, 4, \dodoi{10.3847/1538-4357/aae689}

\bibitem[{Aartsen {et~al.}(2020{\natexlab{a}})Aartsen, Ackermann, Adams,
  Aguilar, Ahlers, Ahrens, Alispach, Andeen, Anderson, Ansseau, Anton,
  Argüelles, Auffenberg, Axani, Backes, Bagherpour, Bai, Balagopal, Barbano,
  Barwick, Bastian, Baum, Baur, Bay, Beatty, Becker, Tjus, BenZvi, Berley,
  Bernardini, Besson, Binder, Bindig, Blaufuss, Blot, Bohm, Börner, Böser,
  Botner, Böttcher, Bourbeau, Bourbeau, Bradascio, Braun, Bron,
  Brostean-Kaiser, Burgman, Buscher, Busse, Carver, Chen, Cheung, Chirkin,
  Choi, Clark, Classen, Coleman, Collin, Conrad, Coppin, Correa, Cowen, Cross,
  Dave, Clercq, DeLaunay, Dembinski, Deoskar, Ridder, Desiati, de~Vries,
  de~Wasseige, de~With, DeYoung, Diaz, D{\'{\i}}az-V{\'{e}}lez, Dujmovic,
  Dunkman, Dvorak, Eberhardt, Ehrhardt, Eller, Engel, Evenson, Fahey, Fazely,
  Felde, Filimonov, Finley, Fox, Franckowiak, Friedman, Fritz, Gaisser,
  Gallagher, Ganster, Garrappa, Gerhardt, Ghorbani, Glauch, Glüsenkamp,
  Goldschmidt, Gonzalez, Grant, Griffith, Griswold, Günder, Gündüz, Haack,
  Hallgren, Halliday, Halve, Halzen, Hanson, Haungs, Hebecker, Heereman, Heix,
  Helbing, Hellauer, Henningsen, Hickford, Hignight, Hill, Hoffman, Hoffmann,
  Hoinka, Hokanson-Fasig, Hoshina, Huang, Huber, Huber, Hultqvist, Hünnefeld,
  Hussain, In, Iovine, Ishihara, Japaridze, Jeong, Jero, Jones, Jonske, Joppe,
  Kang, Kang, Kappes, Kappesser, Karg, Karl, Karle, Katz, Kauer, Kelley,
  Kheirandish, Kim, Kintscher, Kiryluk, Kittler, Klein, Koirala, Kolanoski,
  Köpke, Kopper, Kopper, Koskinen, Kowalski, Krings, Krückl, Kulacz,
  Kurahashi, Kyriacou, Labare, Lanfranchi, Larson, Lauber, Lazar, Leonard,
  Leszczy{\'{n}}ska, Leuermann, Liu, Lohfink, Mariscal, Lu, Lucarelli,
  Lünemann, Luszczak, Lyu, Ma, Madsen, Maggi, Mahn, Makino, Mallik, Mallot,
  Mancina, Mari{\c{s}}, Maruyama, Mase, Matis, Maunu, McNally, Meagher, Medici,
  Medina, Meier, Meighen-Berger, Menne, Merino, Meures, Micallef, Mockler,
  Moment{\'{e}}, Montaruli, Moore, Morse, Moulai, Muth, Nagai, Naumann, Neer,
  Niederhausen, Nisa, Nowicki, Nygren, Pollmann, Oehler, Olivas, O'Murchadha,
  O'Sullivan, Palczewski, Pandya, Pankova, Park, Peiffer, de~los Heros,
  Philippen, Pieloth, Pinat, Pizzuto, Plum, Porcelli, Price, Przybylski, Raab,
  Raissi, Rameez, Rauch, Rawlins, Rea, Reimann, Relethford, Renschler, Renzi,
  Resconi, Rhode, Richman, Robertson, Rongen, Rott, Ruhe, Ryckbosch, Rysewyk,
  Safa, Herrera, Sandrock, Sandroos, Santander, Sarkar, Sarkar, Satalecka,
  Schaufel, Schieler, Schlunder, Schmidt, Schneider, Schneider, Schröder,
  Schumacher, Sclafani, Seckel, Seunarine, Shefali, Silva, Snihur,
  Soedingrekso, Soldin, Song, Spiczak, Spiering, Stachurska, Stamatikos,
  Stanev, Stein, Steinmüller, Stettner, Steuer, Stezelberger, Stokstad,
  Stö{\ss}l, Strotjohann, Stürwald, Stuttard, Sullivan, Taboada, Tenholt,
  Ter-Antonyan, Terliuk, Tilav, Tollefson, Tomankova, Tönnis, Toscano, Tosi,
  Trettin, Tselengidou, Tung, Turcati, Turcotte, Turley, Ty, Unger, Elorrieta,
  Usner, Vandenbroucke, Driessche, van Eijk, van Eijndhoven, Vanheule, van
  Santen, Vraeghe, Walck, Wallace, Wallraff, Wandkowsky, Watson, Weaver,
  Weindl, Weiss, Weldert, Wendt, Werthebach, Whelan, Whitehorn, Wiebe,
  Wiebusch, Wille, Williams, Wills, Wolf, Wood, Wood, Woschnagg, Wrede, Xu, Xu,
  Xu, Yanez, Yodh, Yoshida, Yuan, \& Zöcklein}]{Aartsen_2020}
Aartsen, M., Ackermann, M., Adams, J., {et~al.} 2020{\natexlab{a}}, Physical
  Review Letters, 124, \dodoi{10.1103/physrevlett.124.051103}

\bibitem[{Aartsen {et~al.}(2020{\natexlab{b}})Aartsen, Ackermann, Adams,
  Aguilar, Ahlers, Ahrens, Alispach, Andeen, Anderson, Ansseau, Anton,
  Argüelles, Auffenberg, \& et~al.}]{IceCube_2020}
---. 2020{\natexlab{b}}, Physical Review Letters, 124,
  \dodoi{10.1103/physrevlett.124.051103}

\bibitem[{Abbasi {et~al.}(2017)Abbasi, Abe, Abu-Zayyad, Allen, Azuma,
  Barcikowski, Belz, Bergman, Blake, Cady, Cheon, Chiba, Chikawa, Fujii,
  Fukushima, Goto, Hanlon, Hayashi, Hayashi, Hayashida, Hibino, Honda, Ikeda,
  Inoue, Ishii, Ishimori, Ito, Ivanov, Jui, Kadota, Kakimoto, Kalashev,
  Kasahara, Kawai, Kawakami, Kawana, Kawata, Kido, Kim, Kim, Kim, Kishigami,
  Kitamura, Kitamura, Kuzmin, Kwon, Lan, Lubsandorzhiev, Lundquist, Machida,
  Martens, Matsuda, Matsuyama, Matthews, Minamino, Mukai, Myers, Nagasawa,
  Nagataki, Nakamura, Nonaka, Nozato, Ogio, Ogura, Ohnishi, Ohoka, Oki, Okuda,
  Ono, Onogi, Oshima, Ozawa, Park, Pshirkov, Rodriguez, Rubtsov, Ryu, Sagawa,
  Saito, Saito, Sakaki, Sakurai, Scott, Sekino, Shah, Shibata, Shibata,
  Shimodaira, Shin, Shin, Smith, Sokolsky, Stokes, Stratton, Stroman, Suzawa,
  Takahashi, Takamura, Takeda, Takeishi, Taketa, Takita, Tameda, Tanaka,
  Tanaka, Tanaka, Thomas, Thomson, Tinyakov, Tirone, Tkachev, Tokuno, Tomida,
  Troitsky, Tsunesada, Tsutsumi, Uchihori, Udo, Urban, Wong, Yamane, Yamaoka,
  Yamazaki, Yang, Yashiro, Yoneda, Yoshida, Yoshii, Zollinger, \&
  Zundel}]{Abbasi_2017}
Abbasi, R., Abe, M., Abu-Zayyad, T., {et~al.} 2017, Astroparticle Physics, 86,
  21, \dodoi{10.1016/j.astropartphys.2016.11.001}

\bibitem[{Abbasi {et~al.}(2021)}]{IceCube:2021oqh}
Abbasi, R., {et~al.} 2021, PoS, ICRC2021, 971, \dodoi{10.22323/1.395.0971}

\bibitem[{{Abdo} {et~al.}(2009){Abdo}, {Ackermann}, {Ajello}, {Baldini},
  {Ballet}, {Barbiellini}, {Bastieri}, {Bechtol}, {Bellazzini}, {Berenji},
  {Bloom}, {Bonamente}, {Borgland}, {Bregeon}, {Brez}, {Brigida}, {Bruel},
  {Burnett}, {Caliandro}, {Cameron}, {Caraveo}, {Casandjian}, {Cecchi},
  {{\c{C}}elik}, {Chekhtman}, {Cheung}, {Chiang}, {Ciprini}, {Claus},
  {Cohen-Tanugi}, {Conrad}, {Cutini}, {Dermer}, {de Palma}, {Silva}, {Drell},
  {Dubois}, {Dumora}, {Farnier}, {Favuzzi}, {Fegan}, {Focke}, {Foschini},
  {Frailis}, {Fukazawa}, {Fusco}, {Gargano}, {Gehrels}, {Germani}, {Giebels},
  {Giglietto}, {Giordano}, {Giroletti}, {Glanzman}, {Godfrey}, {Grenier},
  {Grove}, {Guillemot}, {Guiriec}, {Hayashida}, {Hays}, {Horan}, {Hughes},
  {J{\'o}hannesson}, {Johnson}, {Johnson}, {Kadler}, {Kamae}, {Katagiri},
  {Kataoka}, {Kerr}, {Kn{\"o}dlseder}, {Kuss}, {Lande}, {Latronico}, {Longo},
  {Loparco}, {Lott}, {Lovellette}, {Lubrano}, {Makeev}, {Mazziotta},
  {McConville}, {McEnery}, {Meurer}, {Michelson}, {Mitthumsiri}, {Mizuno},
  {Monte}, {Monzani}, {Morselli}, {Moskalenko}, {Murgia}, {Nolan}, {Norris},
  {Nuss}, {Ohsugi}, {Omodei}, {Orlando}, {Ormes}, {Pelassa}, {Pepe}, {Persic},
  {Pesce-Rollins}, {Piron}, {Porter}, {Rain{\`o}}, {Rando}, {Razzano},
  {Rochester}, {Rodriguez}, {Ryde}, {Sadrozinski}, {Sambruna}, {Sander}, {Saz
  Parkinson}, {Scargle}, {Sgr{\`o}}, {Smith}, {Spandre}, {Spinelli},
  {Strickman}, {Suson}, {Tagliaferri}, {Takahashi}, {Takahashi}, {Tanaka},
  {Thayer}, {Thayer}, {Thompson}, {Tibaldo}, {Tibolla}, {Torres}, {Tosti},
  {Tramacere}, {Uchiyama}, {Usher}, {Vasileiou}, {Vilchez}, {Vitale}, {Waite},
  {Wang}, {Winer}, {Wood}, {Ylinen}, {Ziegler}, {Fermi/LAT Collaboration},
  {Ghisellini}, {Maraschi}, \& {Tavecchio}}]{2009ApJ...707L.142A_Abdo}
{Abdo}, A.~A., {Ackermann}, M., {Ajello}, M., {et~al.} 2009, \apjl, 707, L142,
  \dodoi{10.1088/0004-637X/707/2/L142}

\bibitem[{{Abdo} {et~al.}(2010){Abdo}, {Ackermann}, {Agudo}, {Ajello}, {Aller},
  {Aller}, {Angelakis}, {Arkharov}, {Axelsson}, {Bach}, {Baldini}, {Ballet},
  {Barbiellini}, {Bastieri}, {Baughman}, {Bechtol}, {Bellazzini}, {Benitez},
  {Berdyugin}, {Berenji}, {Blandford}, {Bloom}, {Boettcher}, {Bonamente},
  {Borgland}, {Bregeon}, {Brez}, {Brigida}, {Bruel}, {Burnett}, {Burrows},
  {Buson}, {Caliandro}, {Calzoletti}, {Cameron}, {Capalbi}, {Caraveo},
  {Carosati}, {Casandjian}, {Cavazzuti}, {Cecchi}, {{\c{C}}elik}, {Charles},
  {Chaty}, {Chekhtman}, {Chen}, {Chiang}, {Chincarini}, {Ciprini}, {Claus},
  {Cohen-Tanugi}, {Colafrancesco}, {Cominsky}, {Conrad}, {Costamante},
  {Cutini}, {D'ammando}, {Deitrick}, {D'Elia}, {Dermer}, {de Angelis}, {de
  Palma}, {Digel}, {Donnarumma}, {Silva}, {Drell}, {Dubois}, {Dultzin},
  {Dumora}, {Falcone}, {Farnier}, {Favuzzi}, {Fegan}, {Focke}, {Forn{\'e}},
  {Fortin}, {Frailis}, {Fuhrmann}, {Fukazawa}, {Funk}, {Fusco}, {G{\'o}mez},
  {Gargano}, {Gasparrini}, {Gehrels}, {Germani}, {Giebels}, {Giglietto},
  {Giommi}, {Giordano}, {Giuliani}, {Glanzman}, {Godfrey}, {Grenier},
  {Gronwall}, {Grove}, {Guillemot}, {Guiriec}, {Gurwell}, {Hadasch},
  {Hanabata}, {Harding}, {Hayashida}, {Hays}, {Healey}, {Heidt}, {Hiriart},
  {Horan}, {Hoversten}, {Hughes}, {Itoh}, {Jackson}, {J{\'o}hannesson},
  {Johnson}, {Johnson}, {Jorstad}, {Kadler}, {Kamae}, {Katagiri}, {Kataoka},
  {Kawai}, {Kennea}, {Kerr}, {Kimeridze}, {Kn{\"o}dlseder}, {Kocian},
  {Kopatskaya}, {Koptelova}, {Konstantinova}, {Kovalev}, {Kovalev},
  {Kurtanidze}, {Kuss}, {Lande}, {Larionov}, {Latronico}, {Leto}, {Lindfors},
  {Longo}, {Loparco}, {Lott}, {Lovellette}, {Lubrano}, {Madejski}, {Makeev},
  {Marchegiani}, {Marscher}, {Marshall}, {Max-Moerbeck}, {Mazziotta},
  {McConville}, {McEnery}, {Meurer}, {Michelson}, {Mitthumsiri}, {Mizuno},
  {Moiseev}, {Monte}, {Monzani}, {Morselli}, {Moskalenko}, {Murgia},
  {Nestoras}, {Nilsson}, {Nizhelsky}, {Nolan}, {Norris}, {Nuss}, {Ohsugi},
  {Ojha}, {Omodei}, {Orlando}, {Ormes}, {Osborne}, {Ozaki}, {Pacciani},
  {Padovani}, {Pagani}, {Page}, {Paneque}, {Panetta}, {Parent}, {Pasanen},
  {Pavlidou}, {Pelassa}, {Pepe}, {Perri}, {Pesce-Rollins}, {Piranomonte},
  {Piron}, {Pittori}, {Porter}, {Puccetti}, {Rahoui}, {Rain{\`o}}, {Raiteri},
  {Rando}, {Razzano}, {Reimer}, {Reimer}, {Reposeur}, {Richards}, {Ritz},
  {Rochester}, {Rodriguez}, {Romani}, {Ros}, {Roth}, {Roustazadeh}, {Ryde},
  {Sadrozinski}, {Sadun}, {Sanchez}, {Sander}, {Saz Parkinson}, {Scargle},
  {Sellerholm}, {Sgr{\`o}}, {Shaw}, {Sigua}, {Siskind}, {Smith}, {Smith},
  {Spandre}, {Spinelli}, {Starck}, {Stevenson}, {Stratta}, {Strickman},
  {Suson}, {Tajima}, {Takahashi}, {Takahashi}, {Takalo}, {Tanaka}, {Thayer},
  {Thayer}, {Thompson}, {Tibaldo}, {Torres}, {Tosti}, {Tramacere}, {Uchiyama},
  {Usher}, {Vasileiou}, {Verrecchia}, {Vilchez}, {Villata}, {Vitale}, {Waite},
  {Wang}, {Winer}, {Wood}, {Ylinen}, {Zensus}, {Zhekanis}, \&
  {Ziegler}}]{2010ApJ...716...30A}
{Abdo}, A.~A., {Ackermann}, M., {Agudo}, I., {et~al.} 2010, \apj, 716, 30,
  \dodoi{10.1088/0004-637X/716/1/30}

\bibitem[{{Aharonian}(2000)}]{Aharonian_2000}
{Aharonian}, F.~A. 2000, \na, 5, 377, \dodoi{10.1016/S1384-1076(00)00039-7}

\bibitem[{{Alves Batista} {et~al.}(2019){Alves Batista}, {Biteau},
  {Bustamante}, {Dolag}, {Engel}, {Fang}, {Kampert}, {Kostunin}, {Mostafa},
  {Murase}, {Oikonomou}, {Olinto}, {Panasyuk}, {Sigl}, {Taylor}, \&
  {Unger}}]{2019FrASS...6...23B}
{Alves Batista}, R., {Biteau}, J., {Bustamante}, M., {et~al.} 2019, Frontiers
  in Astronomy and Space Sciences, 6, 23, \dodoi{10.3389/fspas.2019.00023}

\bibitem[{{Ansoldi} {et~al.}(2018{\natexlab{a}}){Ansoldi}, {Antonelli},
  {Arcaro}, {Baack}, {Babi{\'c}}, {Banerjee}, {Bangale}, {Barres de Almeida},
  {Barrio}, {Becerra Gonz{\'a}lez}, {Bednarek}, {Bernardini}, {Berse}, {Berti},
  {Besenrieder}, {Bhattacharyya}, {Bigongiari}, {Biland}, {Blanch}, {Bonnoli},
  {Carosi}, {Ceribella}, {Chatterjee}, {Colak}, {Colin}, {Colombo},
  {Contreras}, {Cortina}, {Covino}, {Cumani}, {D'Elia}, {Da Vela}, {Dazzi}, {De
  Angelis}, {De Lotto}, {Delfino}, {Delgado}, {Di Pierro}, {Dom{\'\i}nguez},
  {Dominis Prester}, {Dorner}, {Doro}, {Einecke}, {Elsaesser}, {Fallah
  Ramazani}, {Fattorini}, {Fern{\'a}ndez-Barral}, {Ferrara}, {Fidalgo},
  {Foffano}, {Fonseca}, {Font}, {Fruck}, {Gallozzi}, {Garc{\'\i}a L{\'o}pez},
  {Garczarczyk}, {Gaug}, {Giammaria}, {Godinovi{\'c}}, {Guberman}, {Hadasch},
  {Hahn}, {Hassan}, {Hayashida}, {Herrera}, {Hoang}, {Hrupec}, {Inoue},
  {Ishio}, {Iwamura}, {Konno}, {Kubo}, {Kushida}, {Lamastra}, {Lelas}, {Leone},
  {Lindfors}, {Lombardi}, {Longo}, {L{\'o}pez}, {Maggio}, {Majumdar},
  {Makariev}, {Maneva}, {Manganaro}, {Mannheim}, {Maraschi}, {Mariotti},
  {Mart{\'\i}nez}, {Masuda}, {Mazin}, {Mielke}, {Minev}, {Miranda}, {Mirzoyan},
  {Moralejo}, {Moreno}, {Moretti}, {Neustroev}, {Niedzwiecki}, {Nievas
  Rosillo}, {Nigro}, {Nilsson}, {Ninci}, {Nishijima}, {Noda}, {Nogu{\'e}s},
  {Paiano}, {Palacio}, {Paneque}, {Paoletti}, {Paredes}, {Pedaletti},
  {Pe{\~n}il}, {Peresano}, {Persic}, {Pfrang}, {Prada Moroni}, {Prandini},
  {Puljak}, {Garcia}, {Rhode}, {Rib{\'o}}, {Rico}, {Righi}, {Rugliancich},
  {Saha}, {Saito}, {Satalecka}, {Schweizer}, {Sitarek}, {{\v{S}}nidari{\'c}},
  {Sobczynska}, {Stamerra}, {Strzys}, {Suri{\'c}}, {Tavecchio}, {Temnikov},
  {Terzi{\'c}}, {Teshima}, {Torres-Alb{\'a}}, {Tsujimoto}, {Vanzo}, {Vazquez
  Acosta}, {Vovk}, {Ward}, {Will}, {Zari{\'c}}, \& {Cerruti}}]{Ansoldi_2018}
{Ansoldi}, S., {Antonelli}, L.~A., {Arcaro}, C., {et~al.} 2018{\natexlab{a}},
  \apjl, 863, L10, \dodoi{10.3847/2041-8213/aad083}

\bibitem[{{Ansoldi} {et~al.}(2018{\natexlab{b}}){Ansoldi}, {Antonelli},
  {Arcaro}, {Baack}, {Babi{\'c}}, {Banerjee}, {Bangale}, {Barres de Almeida},
  {Barrio}, {Becerra Gonz{\'a}lez}, {Bednarek}, {Bernardini}, {Berse}, {Berti},
  {Besenrieder}, {Bhattacharyya}, {Bigongiari}, {Biland}, {Blanch}, {Bonnoli},
  {Carosi}, {Ceribella}, {Chatterjee}, {Colak}, {Colin}, {Colombo},
  {Contreras}, {Cortina}, {Covino}, {Cumani}, {D'Elia}, {Da Vela}, {Dazzi}, {De
  Angelis}, {De Lotto}, {Delfino}, {Delgado}, {Di Pierro}, {Dom{\'\i}nguez},
  {Dominis Prester}, {Dorner}, {Doro}, {Einecke}, {Elsaesser}, {Fallah
  Ramazani}, {Fattorini}, {Fern{\'a}ndez-Barral}, {Ferrara}, {Fidalgo},
  {Foffano}, {Fonseca}, {Font}, {Fruck}, {Gallozzi}, {Garc{\'\i}a L{\'o}pez},
  {Garczarczyk}, {Gaug}, {Giammaria}, {Godinovi{\'c}}, {Guberman}, {Hadasch},
  {Hahn}, {Hassan}, {Hayashida}, {Herrera}, {Hoang}, {Hrupec}, {Inoue},
  {Ishio}, {Iwamura}, {Konno}, {Kubo}, {Kushida}, {Lamastra}, {Lelas}, {Leone},
  {Lindfors}, {Lombardi}, {Longo}, {L{\'o}pez}, {Maggio}, {Majumdar},
  {Makariev}, {Maneva}, {Manganaro}, {Mannheim}, {Maraschi}, {Mariotti},
  {Mart{\'\i}nez}, {Masuda}, {Mazin}, {Mielke}, {Minev}, {Miranda}, {Mirzoyan},
  {Moralejo}, {Moreno}, {Moretti}, {Neustroev}, {Niedzwiecki}, {Nievas
  Rosillo}, {Nigro}, {Nilsson}, {Ninci}, {Nishijima}, {Noda}, {Nogu{\'e}s},
  {Paiano}, {Palacio}, {Paneque}, {Paoletti}, {Paredes}, {Pedaletti},
  {Pe{\~n}il}, {Peresano}, {Persic}, {Pfrang}, {Prada Moroni}, {Prandini},
  {Puljak}, {Garcia}, {Rhode}, {Rib{\'o}}, {Rico}, {Righi}, {Rugliancich},
  {Saha}, {Saito}, {Satalecka}, {Schweizer}, {Sitarek}, {{\v{S}}nidari{\'c}},
  {Sobczynska}, {Stamerra}, {Strzys}, {Suri{\'c}}, {Tavecchio}, {Temnikov},
  {Terzi{\'c}}, {Teshima}, {Torres-Alb{\'a}}, {Tsujimoto}, {Vanzo}, {Vazquez
  Acosta}, {Vovk}, {Ward}, {Will}, {Zari{\'c}}, \&
  {Cerruti}}]{2018ApJ...863L..10A}
---. 2018{\natexlab{b}}, \apjl, 863, L10, \dodoi{10.3847/2041-8213/aad083}

\bibitem[{{Atoyan} \& {Dermer}(2001)}]{2001PhRvL..87v1102A}
{Atoyan}, A., \& {Dermer}, C.~D. 2001, \prl, 87, 221102,
  \dodoi{10.1103/PhysRevLett.87.221102}

\bibitem[{{Banik} \& {Bhadra}(2019)}]{2019PhRvD..99j3006B}
{Banik}, P., \& {Bhadra}, A. 2019, \prd, 99, 103006,
  \dodoi{10.1103/PhysRevD.99.103006}

\bibitem[{Böttcher {et~al.}(2013)Böttcher, Reimer, Sweeney, \&
  Prakash}]{2013_Boettcher}
Böttcher, M., Reimer, A., Sweeney, K., \& Prakash, A. 2013, The Astrophysical
  Journal, 768, 54, \dodoi{10.1088/0004-637x/768/1/54}

\bibitem[{Celotti \& Ghisellini(2008)}]{Celotti_2008}
Celotti, A., \& Ghisellini, G. 2008, Monthly Notices of the Royal Astronomical
  Society, 385, 283, \dodoi{10.1111/j.1365-2966.2007.12758.x}

\bibitem[{Cerruti(2020)}]{Cerruti_2020}
Cerruti, M. 2020, Galaxies, 8, 72, \dodoi{10.3390/galaxies8040072}

\bibitem[{{Cerruti} {et~al.}(2019){Cerruti}, {Zech}, {Boisson}, {Emery},
  {Inoue}, \& {Lenain}}]{2019MNRAS.483L..12C}
{Cerruti}, M., {Zech}, A., {Boisson}, C., {et~al.} 2019, \mnras, 483, L12,
  \dodoi{10.1093/mnrasl/sly210}

\bibitem[{{Cleary} {et~al.}(2007){Cleary}, {Lawrence}, {Marshall}, {Hao}, \&
  {Meier}}]{2007ApJ...660..117C}
{Cleary}, K., {Lawrence}, C.~R., {Marshall}, J.~A., {Hao}, L., \& {Meier}, D.
  2007, \apj, 660, 117, \dodoi{10.1086/511969}

\bibitem[{D{\textquotesingle}Ammando {et~al.}(2013)D{\textquotesingle}Ammando,
  Orienti, Doi, Giroletti, Dallacasa, Hovatta, Drake, Max-Moerbeck, Readhead,
  \& Richards}]{D_Ammando_2013}
D{\textquotesingle}Ammando, F., Orienti, M., Doi, A., {et~al.} 2013, Monthly
  Notices of the Royal Astronomical Society, 433, 952,
  \dodoi{10.1093/mnras/stt778}

\bibitem[{Doi {et~al.}(2006)Doi, Nagai, Asada, Kameno, Wajima, \&
  Inoue}]{Doi_2006}
Doi, A., Nagai, H., Asada, K., {et~al.} 2006, Publ. Astron. Soc. Jap., 58, 829,
  \dodoi{10.1093/pasj/58.5.829}

\bibitem[{Foreman-Mackey {et~al.}(2013)Foreman-Mackey, Hogg, Lang, \&
  Goodman}]{emcee_2013}
Foreman-Mackey, D., Hogg, D.~W., Lang, D., \& Goodman, J. 2013, Publications of
  the Astronomical Society of the Pacific, 125, 306–312,
  \dodoi{10.1086/670067}

\bibitem[{{Fossati} {et~al.}(1998){Fossati}, {Maraschi}, {Celotti}, {Comastri},
  \& {Ghisellini}}]{1998MNRAS.299..433F}
{Fossati}, G., {Maraschi}, L., {Celotti}, A., {Comastri}, A., \& {Ghisellini},
  G. 1998, \mnras, 299, 433, \dodoi{10.1046/j.1365-8711.1998.01828.x}

\bibitem[{{Gao} {et~al.}(2019){Gao}, {Fedynitch}, {Winter}, \&
  {Pohl}}]{2019NatAs...3...88G}
{Gao}, S., {Fedynitch}, A., {Winter}, W., \& {Pohl}, M. 2019, Nature Astronomy,
  3, 88, \dodoi{10.1038/s41550-018-0610-1}

\bibitem[{{Ghisellini} {et~al.}(2010){Ghisellini}, {Tavecchio}, {Foschini},
  {Ghirlanda}, {Maraschi}, \& {Celotti}}]{2010MNRAS.402..497G}
{Ghisellini}, G., {Tavecchio}, F., {Foschini}, L., {et~al.} 2010, \mnras, 402,
  497, \dodoi{10.1111/j.1365-2966.2009.15898.x}

\bibitem[{{Giommi} {et~al.}(1995){Giommi}, {Ansari}, \&
  {Micol}}]{1995A&AS..109..267G}
{Giommi}, P., {Ansari}, S.~G., \& {Micol}, A. 1995, \aaps, 109, 267

\bibitem[{Halzen \& Hooper(2002)}]{Halzen_2002}
Halzen, F., \& Hooper, D. 2002, Reports on Progress in Physics, 65, 1025,
  \dodoi{10.1088/0034-4885/65/7/201}

\bibitem[{Harris \& Krawczynski(2006)}]{Harris_2006}
Harris, D., \& Krawczynski, H. 2006, Annual Review of Astronomy and
  Astrophysics, 44, 463–506, \dodoi{10.1146/annurev.astro.44.051905.092446}

\bibitem[{{Hayashida} {et~al.}(2012){Hayashida}, {Madejski}, {Nalewajko},
  {Sikora}, {Wehrle}, {Ogle}, {Collmar}, {Larsson}, {Fukazawa}, {Itoh},
  {Chiang}, {Stawarz}, {Blandford}, {Richards}, {Max-Moerbeck}, {Readhead},
  {Buehler}, {Cavazzuti}, {Ciprini}, {Gehrels}, {Reimer}, {Szostek}, {Tanaka},
  {Tosti}, {Uchiyama}, {Kawabata}, {Kino}, {Sakimoto}, {Sasada}, {Sato},
  {Uemura}, {Yamanaka}, {Greiner}, {Kruehler}, {Rossi}, {Macquart}, {Bock},
  {Villata}, {Raiteri}, {Agudo}, {Aller}, {Aller}, {Arkharov}, {Bach},
  {Ben{\'\i}tez}, {Berdyugin}, {Blinov}, {Blumenthal}, {B{\"o}ttcher}, {Buemi},
  {Carosati}, {Chen}, {Di Paola}, {Dolci}, {Efimova}, {Forn{\'e}}, {G{\'o}mez},
  {Gurwell}, {Heidt}, {Hiriart}, {Jordan}, {Jorstad}, {Joshi}, {Kimeridze},
  {Konstantinova}, {Kopatskaya}, {Koptelova}, {Kurtanidze},
  {L{\"a}hteenm{\"a}ki}, {Lamerato}, {Larionov}, {Larionova}, {Larionova},
  {Leto}, {Lindfors}, {Marscher}, {McHardy}, {Molina}, {Morozova},
  {Nikolashvili}, {Nilsson}, {Reinthal}, {Roustazadeh}, {Sakamoto}, {Sigua},
  {Sillanp{\"a}{\"a}}, {Takalo}, {Tammi}, {Taylor}, {Tornikoski}, {Trigilio},
  {Troitsky}, \& {Umana}}]{2012ApJ...754..114H}
{Hayashida}, M., {Madejski}, G.~M., {Nalewajko}, K., {et~al.} 2012, \apj, 754,
  114, \dodoi{10.1088/0004-637X/754/2/114}

\bibitem[{{Hillas}(1984)}]{1984ARA&A..22..425H}
{Hillas}, A.~M. 1984, \araa, 22, 425,
  \dodoi{10.1146/annurev.aa.22.090184.002233}

\bibitem[{{Huchra} \& {Burg}(1992)}]{1992ApJ...393...90H}
{Huchra}, J., \& {Burg}, R. 1992, \apj, 393, 90, \dodoi{10.1086/171488}

\bibitem[{{IceCube Collaboration} {et~al.}(2018{\natexlab{a}}){IceCube
  Collaboration}, {Aartsen}, {Ackermann}, {Adams}, {Aguilar}, {Ahlers},
  {Ahrens}, {Al Samarai}, {Altmann}, {Andeen}, {Anderson}, {Ansseau}, {Anton},
  {Arg{\"u}elles}, {Auffenberg}, {Axani}, {Bagherpour}, {Bai}, {Barron},
  {Barwick}, {Baum}, {Bay}, {Beatty}, {Becker Tjus}, {Becker}, {BenZvi},
  {Berley}, {Bernardini}, {Besson}, {Binder}, {Bindig}, {Blaufuss}, {Blot},
  {Bohm}, {B{\"o}rner}, {Bos}, {B{\"o}ser}, {Botner}, {Bourbeau}, {Bourbeau},
  {Bradascio}, {Braun}, {Brenzke}, {Bretz}, {Bron}, {Brostean-Kaiser},
  {Burgman}, {Busse}, {Carver}, {Cheung}, {Chirkin}, {Christov}, {Clark},
  {Classen}, {Coenders}, {Collin}, {Conrad}, {Coppin}, {Correa}, {Cowen},
  {Cross}, {Dave}, {Day}, {de Andr{\'e}}, {De Clercq}, {DeLaunay}, {Dembinski},
  {De Ridder}, {Desiati}, {de Vries}, {de Wasseige}, {de With}, {DeYoung},
  {D{\'\i}az-V{\'e}lez}, {di Lorenzo}, {Dujmovic}, {Dumm}, {Dunkman}, {Dvorak},
  {Eberhardt}, {Ehrhardt}, {Eichmann}, {Eller}, {Evenson}, {Fahey}, {Fazely},
  {Felde}, {Filimonov}, {Finley}, {Flis}, {Franckowiak}, {Friedman}, {Fritz},
  {Gaisser}, {Gallagher}, {Gerhardt}, {Ghorbani}, {Glauch}, {Gl{\"u}senkamp},
  {Goldschmidt}, {Gonzalez}, {Grant}, {Griffith}, {Haack}, {Hallgren},
  {Halzen}, {Hanson}, {Hebecker}, {Heereman}, {Helbing}, {Hellauer},
  {Hickford}, {Hignight}, {Hill}, {Hoffman}, {Hoffmann}, {Hoinka},
  {Hokanson-Fasig}, {Hoshina}, {Huang}, {Huber}, {Hultqvist}, {H{\"u}nnefeld},
  {Hussain}, {In}, {Iovine}, {Ishihara}, {Jacobi}, {Japaridze}, {Jeong},
  {Jero}, {Jones}, {Kalaczynski}, {Kang}, {Kappes}, {Kappesser}, {Karg},
  {Karle}, {Katz}, {Kauer}, {Keivani}, {Kelley}, {Kheirandish}, {Kim}, {Kim},
  {Kintscher}, {Kiryluk}, {Kittler}, {Klein}, {Koirala}, {Kolanoski},
  {K{\"o}pke}, {Kopper}, {Kopper}, {Koschinsky}, {Koskinen}, {Kowalski},
  {Krings}, {Kroll}, {Kr{\"u}ckl}, {Kunwar}, {Kurahashi}, {Kuwabara},
  {Kyriacou}, {Labare}, {Lanfranchi}, {Larson}, {Lauber}, {Leonard},
  {Lesiak-Bzdak}, {Leuermann}, {Liu}, {Lozano Mariscal}, {Lu}, {L{\"u}nemann},
  {Luszczak}, {Madsen}, {Maggi}, {Mahn}, {Mancina}, {Maruyama}, {Mase},
  {Maunu}, {Meagher}, {Medici}, {Meier}, {Menne}, {Merino}, {Meures},
  {Miarecki}, {Micallef}, {Moment{\'e}}, {Montaruli}, {Moore}, {Morse},
  {Moulai}, {Nahnhauer}, {Nakarmi}, {Naumann}, {Neer}, {Niederhausen},
  {Nowicki}, {Nygren}, {Obertacke Pollmann}, {Olivas}, {O'Murchadha},
  {O'Sullivan}, {Palczewski}, {Pandya}, {Pankova}, {Peiffer}, {Pepper},
  {P{\'e}rez de los Heros}, {Pieloth}, {Pinat}, {Plum}, {Price}, {Przybylski},
  {Raab}, {R{\"a}del}, {Rameez}, {Rauch}, {Rawlins}, {Rea}, {Reimann},
  {Relethford}, {Relich}, {Resconi}, {Rhode}, {Richman}, {Robertson}, {Rongen},
  {Rott}, {Ruhe}, {Ryckbosch}, {Rysewyk}, {Safa}, {S{\"a}lzer}, {Sanchez
  Herrera}, {Sandrock}, {Sandroos}, {Santander}, {Sarkar}, {Sarkar},
  {Satalecka}, {Schlunder}, {Schmidt}, {Schneider}, {Schoenen},
  {Sch{\"o}neberg}, {Schumacher}, {Sclafani}, {Seckel}, {Seunarine},
  {Soedingrekso}, {Soldin}, {Song}, {Spiczak}, {Spiering}, {Stachurska},
  {Stamatikos}, {Stanev}, {Stasik}, {Stein}, {Stettner}, {Steuer},
  {Stezelberger}, {Stokstad}, {St{\"o}{\ss}l}, {Strotjohann}, {Stuttard},
  {Sullivan}, {Sutherland}, {Taboada}, {Tatar}, {Tenholt}, {Ter-Antonyan},
  {Terliuk}, {Tilav}, {Toale}, {Tobin}, {Toennis}, {Toscano}, {Tosi},
  {Tselengidou}, {Tung}, {Turcati}, {Turley}, {Ty}, {Unger}, {Usner},
  {Vandenbroucke}, {Van Driessche}, {van Eijk}, {van Eijndhoven}, {Vanheule},
  {van Santen}, {Vogel}, {Vraeghe}, {Walck}, {Wallace}, {Wallraff}, {Wandler},
  {Wandkowsky}, {Waza}, {Weaver}, {Weiss}, {Wendt}, {Werthebach}, {Westerhoff},
  {Whelan}, {Whitehorn}, {Wiebe}, {Wiebusch}, {Wille}, {Williams}, {Wills},
  {Wolf}, {Wood}, {Wood}, {Woschnagg}, {Xu}, {Xu}, {Xu}, {Yanez}, {Yodh},
  {Yoshida}, {Yuan}, {Fermi-LAT Collaboration}, {Abdollahi}, {Ajello},
  {Angioni}, {Baldini}, {Ballet}, {Barbiellini}, {Bastieri}, {Bechtol},
  {Bellazzini}, {Berenji}, {Bissaldi}, {Blandford}, {Bonino}, {Bottacini},
  {Bregeon}, {Bruel}, {Buehler}, {Burnett}, {Burns}, {Buson}, {Cameron},
  {Caputo}, {Caraveo}, {Cavazzuti}, {Charles}, {Chen}, {Cheung}, {Chiang},
  {Chiaro}, {Ciprini}, {Cohen-Tanugi}, {Conrad}, {Costantin}, {Cutini},
  {D'Ammando}, {de Palma}, {Digel}, {Di Lalla}, {Di Mauro}, {Di Venere},
  {Dom{\'\i}nguez}, {Favuzzi}, {Franckowiak}, {Fukazawa}, {Funk}, {Fusco},
  {Gargano}, {Gasparrini}, {Giglietto}, {Giomi}, {Giommi}, {Giordano},
  {Giroletti}, {Glanzman}, {Green}, {Grenier}, {Grondin}, {Guiriec}, {Harding},
  {Hayashida}, {Hays}, {Hewitt}, {Horan}, {J{\'o}hannesson}, {Kadler},
  {Kensei}, {Kocevski}, {Krauss}, {Kreter}, {Kuss}, {La Mura}, {Larsson},
  {Latronico}, {Lemoine-Goumard}, {Li}, {Longo}, {Loparco}, {Lovellette},
  {Lubrano}, {Magill}, {Maldera}, {Malyshev}, {Manfreda}, {Mazziotta},
  {McEnery}, {Meyer}, {Michelson}, {Mizuno}, {Monzani}, {Morselli},
  {Moskalenko}, {Negro}, {Nuss}, {Ojha}, {Omodei}, {Orienti}, {Orlando},
  {Palatiello}, {Paliya}, {Perkins}, {Persic}, {Pesce-Rollins}, {Piron},
  {Porter}, {Principe}, {Rain{\`o}}, {Rando}, {Rani}, {Razzano}, {Razzaque},
  {Reimer}, {Reimer}, {Renault-Tinacci}, {Ritz}, {Rochester}, {Saz Parkinson},
  {Sgr{\`o}}, {Siskind}, {Spandre}, {Spinelli}, {Suson}, {Tajima}, {Takahashi},
  {Tanaka}, {Thayer}, {Thompson}, {Tibaldo}, {Torres}, {Torresi}, {Tosti},
  {Troja}, {Valverde}, {Vianello}, {Vogel}, {Wood}, {Wood}, {Zaharijas}, {MAGIC
  Collaboration}, {Ahnen}, {Ansoldi}, {Antonelli}, {Arcaro}, {Baack},
  {Babi{\'c}}, {Banerjee}, {Bangale}, {Barres de Almeida}, {Barrio}, {Becerra
  Gonz{\'a}lez}, {Bednarek}, {Bernardini}, {Berti}, {Bhattacharyya}, {Biland},
  {Blanch}, {Bonnoli}, {Carosi}, {Carosi}, {Ceribella}, {Chatterjee}, {Colak},
  {Colin}, {Colombo}, {Contreras}, {Cortina}, {Covino}, {Cumani}, {Da Vela},
  {Dazzi}, {De Angelis}, {De Lotto}, {Delfino}, {Delgado}, {Di Pierro},
  {Dom{\'\i}nguez}, {Dominis Prester}, {Dorner}, {Doro}, {Einecke},
  {Elsaesser}, {Fallah Ramazani}, {Fern{\'a}ndez-Barral}, {Fidalgo}, {Foffano},
  {Pfrang}, {Fonseca}, {Font}, {Franceschini}, {Fruck}, {Galindo}, {Gallozzi},
  {Garc{\'\i}a L{\'o}pez}, {Garczarczyk}, {Gaug}, {Giammaria}, {Godinovi{\'c}},
  {Gora}, {Guberman}, {Hadasch}, {Hahn}, {Hassan}, {Hayashida}, {Herrera},
  {Hose}, {Hrupec}, {Inoue}, {Ishio}, {Konno}, {Kubo}, {Kushida}, {Lelas},
  {Lindfors}, {Lombardi}, {Longo}, {L{\'o}pez}, {Maggio}, {Majumdar},
  {Makariev}, {Maneva}, {Manganaro}, {Mannheim}, {Maraschi}, {Mariotti},
  {Mart{\'\i}nez}, {Masuda}, {Mazin}, {Minev}, {M}, {Mirzoyan}, {Moralejo},
  {Moreno}, {Moretti}, {Nagayoshi}, {Neustroev}, {Niedzwiecki}, {Nievas
  Rosillo}, {Nigro}, {Nilsson}, {Ninci}, {Nishijima}, {Noda}, {Nogu{\'e}s},
  {Paiano}, {Palacio}, {Paneque}, {Paoletti}, {Paredes}, {Pedaletti},
  {Peresano}, {Persic}, {Prada Moroni}, {Prandini}, {Puljak}, {Rodriguez
  Garcia}, {Reichardt}, {Rhode}, {Rib{\'o}}, {Rico}, {Righi}, {Rugliancich},
  {Saito}, {Satalecka}, {Schweizer}, {Sitarek}, {{\v{S}}nidaric
  {\textasciiacute}}, {Sobczynska}, {Stamerra}, {Strzys}, {Suri{\'c}},
  {Takahashi}, {Tavecchio}, {Temnikov}, {Terzi{\'c}}, {Teshima},
  {Torres-Alb{\`a}}, {Treves}, {Tsujimoto}, {Vanzo}, {Vazquez Acosta}, {Vovk},
  {Ward}, {Will}, {S}, {Zaric {\textasciiacute}}, {AGILE Team}, {Lucarelli},
  {Tavani}, {Piano}, {Donnarumma}, {Pittori}, {Verrecchia}, {Barbiellini},
  {Bulgarelli}, {Caraveo}, {Cattaneo}, {Colafrancesco}, {Costa}, {Di Cocco},
  {Ferrari}, {Gianotti}, {Giuliani}, {Lipari}, {Mereghetti}, {Morselli},
  {Pacciani}, {Paoletti}, {Parmiggiani}, {Pellizzoni}, {Picozza}, {Pilia},
  {Rappoldi}, {Trois}, {Vercellone}, {Vittorini}, {ASAS-SN Team}, {Stanek},
  {Franckowiak}, {Kochanek}, {Beacom}, {Thompson}, {Holoien}, {Dong}, {Prieto},
  {Shappee}, {Holmbo}, {HAWC Collaboration}, {Abeysekara}, {Albert}, {Alfaro},
  {Alvarez}, {Arceo}, {Arteaga-Vel{\'a}zquez}, {Avila Rojas}, {Ayala Solares},
  {Becerril}, {Belmont-Moreno}, {Bernal}, {Caballero-Mora}, {Capistr{\'a}n},
  {Carrami{\~n}ana}, {Casanova}, {Castillo}, {Cotti}, {Cotzomi}, {Couti{\~n}o
  de Le{\'o}n}, {De Le{\'o}n}, {De la Fuente}, {Diaz Hernandez}, {Dichiara},
  {Dingus}, {DuVernois}, {D{\'\i}az-V{\'e}lez}, {Ellsworth}, {Engel},
  {Fiorino}, {Fleischhack}, {Fraija}, {Garc{\'\i}a-Gonz{\'a}lez}, {Garfias},
  {Gonz{\'a}lez Mu{\~n}oz}, {Gonz{\'a}lez}, {Goodman}, {Hampel-Arias},
  {Harding}, {Hernandez}, {Hona}, {Hueyotl-Zahuantitla}, {Hui},
  {H{\"u}ntemeyer}, {Iriarte}, {Jardin-Blicq}, {Joshi}, {Kaufmann}, {Kunde},
  {Lara}, {Lauer}, {Lee}, {Lennarz}, {Le{\'o}n Vargas}, {Linnemann},
  {Longinotti}, {Luis-Raya}, {Luna-Garc{\'\i}a}, {Malone}, {Marinelli},
  {Martinez}, {Martinez-Castellanos}, {Mart{\'\i}nez-Castro},
  {Mart{\'\i}nez-Huerta}, {Matthews}, {Miranda-Romagnoli}, {Moreno},
  {Mostaf{\'a}}, {Nayerhoda}, {Nellen}, {Newbold}, {Nisa}, {Noriega-Papaqui},
  {Pelayo}, {Pretz}, {P{\'e}rez-P{\'e}rez}, {Ren}, {Rho}, {Rivi{\`e}re},
  {Rosa-Gonz{\'a}lez}, {Rosenberg}, {Ruiz-Velasco}, {Ruiz-Velasco}, {Salesa
  Greus}, {Sandoval}, {Schneider}, {Schoorlemmer}, {Sinnis}, {Smith},
  {Springer}, {Surajbali}, {Tibolla}, {Tollefson}, {Torres}, {Villase{\~n}or},
  {Weisgarber}, {Werner}, {Yapici}, {Gaurang}, {Zepeda}, {Zhou}, {{\'A}lvarez},
  {H.~E.~S.~S. Collaboration}, {Abdalla}, {Ang{\"u}ner}, {Armand}, {Backes},
  {Becherini}, {Berge}, {B{\"o}ttcher}, {Boisson}, {Bolmont}, {Bonnefoy},
  {Bordas}, {Brun}, {B{\"u}chele}, {Bulik}, {Caroff}, {Carosi}, {Casanova},
  {Cerruti}, {Chakraborty}, {Chandra}, {Chen}, {Colafrancesco}, {Davids},
  {Deil}, {Devin}, {Djannati-Ata{\"\i}}, {Egberts}, {Emery}, {Eschbach},
  {Fiasson}, {Fontaine}, {Funk}, {F{\"u}{\ss}ling}, {Gallant}, {Gat{\'e}},
  {Giavitto}, {Glawion}, {Glicenstein}, {Gottschall}, {Grondin}, {Haupt},
  {Henri}, {Hinton}, {Hoischen}, {Holch}, {Huber}, {Jamrozy}, {Jankowsky},
  {Jankowsky}, {Jouvin}, {Jung-Richardt}, {Kerszberg}, {Kh{\'e}lifi}, {King},
  {Klepser}, {Kluz {\textasciiacute}niak}, {Komin}, {Kraus}, {Lefaucheur},
  {Lemi{\`e}re}, {Lemoine-Goumard}, {Lenain}, {Leser}, {Lohse},
  {L{\'o}pez-Coto}, {Lorentz}, {Lypova}, {Marandon}, {Guillem
  Mart{\'\i}-Devesa}, {Maurin}, {Mitchell}, {Moderski}, {Mohamed}, {Mohrmann},
  {Moulin}, {Murach}, {de Naurois}, {Niederwanger}, {Niemiec}, {Oakes},
  {O'Brien}, {Ohm}, {Ostrowski}, {Oya}, {Panter}, {Parsons}, {Perennes},
  {Piel}, {Pita}, {Poireau}, {Priyana Noel}, {Prokoph}, {P{\"u}hlhofer},
  {Quirrenbach}, {Raab}, {Rauth}, {Renaud}, {Rieger}, {Rinchiuso}, {Romoli},
  {Rowell}, {Rudak}, {Sasaki}, {Sanchez}, {Schlickeiser}, {Sch{\"u}ssler},
  {Schulz}, {Schwanke}, {Seglar-Arroyo}, {Shafi}, {Simoni}, {Sol}, {Stegmann},
  {Steppa}, {Tavernier}, {Taylor}, {Tiziani}, {Trichard}, {Tsirou}, {van
  Eldik}, {van Rensburg}, {van Soelen}, {Veh}, {Vincent}, {Voisin}, {Wagner},
  {Wagner}, {Wierzcholska}, {Zanin}, {Zdziarski}, {Zech}, {Ziegler}, {Zorn},
  {{\.Z}ywucka}, {INTEGRAL Team}, {Savchenko}, {Ferrigno}, {Bazzano}, {Diehl},
  {Kuulkers}, {Laurent}, {Mereghetti}, {Natalucci}, {Panessa}, {Rodi},
  {Ubertini}, {Kanata}, Teams, {Morokuma}, {Ohta}, {Tanaka}, {Mori},
  {Yamanaka}, {Kawabata}, {Utsumi}, {Nakaoka}, {Kawabata}, {Nagashima},
  {Yoshida}, {Matsuoka}, {Itoh}, {Kapteyn Team}, {Keel}, {Liverpool Telescope
  Team}, {Copperwheat}, {Steele}, {Swift/NuSTAR Team}, {Cenko}, {Cowen},
  {DeLaunay}, {Evans}, {Fox}, {Keivani}, {Kennea}, {Marshall}, {Osborne},
  {Santander}, {Tohuvavohu}, {Turley}, {VERITAS Collaboration}, {Abeysekara},
  {Archer}, {Benbow}, {Bird}, {Brill}, {Brose}, {Buchovecky}, {Buckley},
  {Bugaev}, {Christiansen}, {Connolly}, {Cui}, {Daniel}, {Errando}, {Falcone},
  {Feng}, {Finley}, {Fortson}, {Furniss}, {Gueta}, {H{\"u}tten}, {Hervet},
  {Hughes}, {Humensky}, {Johnson}, {Kaaret}, {Kar}, {Kelley-Hoskins},
  {Kertzman}, {Kieda}, {Krause}, {Krennrich}, {Kumar}, {Lang}, {Lin}, {Maier},
  {McArthur}, {Moriarty}, {Mukherjee}, {Nieto}, {O'Brien}, {Ong}, {Otte},
  {Park}, {Petrashyk}, {Pohl}, {Popkow}, {Pueschel}, {Quinn}, {Ragan},
  {Reynolds}, {Richards}, {Roache}, {Rulten}, {Sadeh}, {Santander}, {Scott},
  {Sembroski}, {Shahinyan}, {Sushch}, {Tr{\'e}panier}, {Tyler}, {Vassiliev},
  {Wakely}, {Weinstein}, {Wells}, {Wilcox}, {Wilhelm}, {Williams}, {Zitzer},
  {VLA/B Team}, {Tetarenko}, {Kimball}, {Miller-Jones}, \&
  {Sivakoff}}]{2018Sci...361.1378I}
{IceCube Collaboration}, {Aartsen}, M.~G., {Ackermann}, M., {et~al.}
  2018{\natexlab{a}}, Science, 361, eaat1378, \dodoi{10.1126/science.aat1378}

\bibitem[{{IceCube Collaboration} {et~al.}(2018{\natexlab{b}}){IceCube
  Collaboration}, {Aartsen}, {Ackermann}, {Adams}, {Aguilar}, {Ahlers},
  {Ahrens}, {Samarai}, {Altmann}, {Andeen}, {Anderson}, {Ansseau}, {Anton},
  {Arg{\"u}elles}, {Arsioli}, {Auffenberg}, {Axani}, {Bagherpour}, {Bai},
  {Barron}, {Barwick}, {Baum}, {Bay}, {Beatty}, {Becker Tjus}, {Becker},
  {BenZvi}, {Berley}, {Bernardini}, {Besson}, {Binder}, {Bindig}, {Blaufuss},
  {Blot}, {Bohm}, {B{\"o}rner}, {Bos}, {B{\"o}ser}, {Botner}, {Bourbeau},
  {Bourbeau}, {Bradascio}, {Braun}, {Brenzke}, {Bretz}, {Bron},
  {Brostean-Kaiser}, {Burgman}, {Busse}, {Carver}, {Cheung}, {Chirkin},
  {Christov}, {Clark}, {Classen}, {Coenders}, {Collin}, {Conrad}, {Coppin},
  {Correa}, {Cowen}, {Cross}, {Dave}, {Day}, {de Andr{\'e}}, {De Clercq},
  {DeLaunay}, {Dembinski}, {DeRidder}, {Desiati}, {de Vries}, {de Wasseige},
  {de With}, {DeYoung}, {D{\'\i}az-V{\'e}lez}, {di Lorenzo}, {Dujmovic},
  {Dumm}, {Dunkman}, {Dvorak}, {Eberhardt}, {Ehrhardt}, {Eichmann}, {Eller},
  {Evenson}, {Fahey}, {Fazely}, {Felde}, {Filimonov}, {Finley}, {Flis},
  {Franckowiak}, {Friedman}, {Fritz}, {Gaisser}, {Gallagher}, {Gerhardt},
  {Ghorbani}, {Giommi}, {Glauch}, {Gl{\"u}senkamp}, {Goldschmidt}, {Gonzalez},
  {Grant}, {Griffith}, {Haack}, {Hallgren}, {Halzen}, {Hanson}, {Hebecker},
  {Heereman}, {Helbing}, {Hellauer}, {Hickford}, {Hignight}, {Hill}, {Hoffman},
  {Hoffmann}, {Hoinka}, {Hokanson-Fasig}, {Hoshina}, {Huang}, {Huber},
  {Hultqvist}, {H{\"u}nnefeld}, {Hussain}, {In}, {Iovine}, {Ishihara},
  {Jacobi}, {Japaridze}, {Jeong}, {Jero}, {Jones}, {Kalaczynski}, {Kang},
  {Kappes}, {Kappesser}, {Karg}, {Karle}, {Katz}, {Kauer}, {Keivani}, {Kelley},
  {Kheirandish}, {Kim}, {Kim}, {Kintscher}, {Kiryluk}, {Kittler}, {Klein},
  {Koirala}, {Kolanoski}, {K{\"o}pke}, {Kopper}, {Kopper}, {Koschinsky},
  {Koskinen}, {Kowalski}, {Krammer}, {Krings}, {Kroll}, {Kr{\"u}ckl}, {Kunwar},
  {Kurahashi}, {Kuwabara}, {Kyriacou}, {Labare}, {Lanfranchi}, {Larson},
  {Lauber}, {Leonard}, {Lesiak-Bzdak}, {Leuermann}, {Liu}, {Lozano Mariscal},
  {Lu}, {L{\"u}nemann}, {Luszczak}, {Madsen}, {Maggi}, {Mahn}, {Mancina},
  {Maruyama}, {Mase}, {Maunu}, {Meagher}, {Medici}, {Meier}, {Menne}, {Merino},
  {Meures}, {Miarecki}, {Micallef}, {Moment{\'e}}, {Montaruli}, {Moore},
  {Morse}, {Moulai}, {Nahnhauer}, {Nakarmi}, {Naumann}, {Neer}, {Niederhausen},
  {Nowicki}, {Nygren}, {Obertacke Pollmann}, {Olivas}, {O'Murchadha},
  {O'Sullivan}, {Padovani}, {Palczewski}, {Pandya}, {Pankova}, {Peiffer},
  {Pepper}, {P{\'e}rez de los Heros}, {Pieloth}, {Pinat}, {Plum}, {Price},
  {Przybylski}, {Raab}, {R{\"a}del}, {Rameez}, {Rawlins}, {Rea}, {Reimann},
  {Relethford}, {Relich}, {Resconi}, {Rhode}, {Richman}, {Robertson}, {Rongen},
  {Rott}, {Ruhe}, {Ryckbosch}, {Rysewyk}, {Safa}, {Sahakyan}, {S{\"a}lzer},
  {Sanchez Herrera}, {Sandrock}, {Sandroos}, {Santander}, {Sarkar}, {Sarkar},
  {Satalecka}, {Schlunder}, {Schmidt}, {Schneider}, {Schoenen},
  {Sch{\"o}neberg}, {Schumacher}, {Sclafani}, {Seckel}, {Seunarine},
  {Soedingrekso}, {Soldin}, {Song}, {Spiczak}, {Spiering}, {Stachurska},
  {Stamatikos}, {Stanev}, {Stasik}, {Stettner}, {Steuer}, {Stezelberger},
  {Stokstad}, {St{\"o}{\ss}l}, {Strotjohann}, {Stuttard}, {Sullivan},
  {Sutherland}, {Taboada}, {Tatar}, {Tenholt}, {Ter-Antonyan}, {Terliuk},
  {Tilav}, {Toale}, {Tobin}, {Toennis}, {Toscano}, {Tosi}, {Tselengidou},
  {Tung}, {Turcati}, {Turley}, {Ty}, {Unger}, {Usner}, {Vandenbroucke}, {Van
  Driessche}, {van Eijk}, {van Eijndhoven}, {Vanheule}, {van Santen}, {Vogel},
  {Vraeghe}, {Walck}, {Wallace}, {Wallraff}, {Wandler}, {Wandkowsky}, {Waza},
  {Weaver}, {Weiss}, {Wendt}, {Werthebach}, {Westerhoff}, {Whelan},
  {Whitehorn}, {Wiebe}, {Wiebusch}, {Wille}, {Williams}, {Wills}, {Wolf},
  {Wood}, {Wood}, {Woschnagg}, {Xu}, {Xu}, {Xu}, {Yanez}, {Yodh}, {Yoshida}, \&
  {Yuan}}]{2018Sci...361..147I}
---. 2018{\natexlab{b}}, Science, 361, 147, \dodoi{10.1126/science.aat2890}

\bibitem[{{Katarzy{\'n}ski} {et~al.}(2001){Katarzy{\'n}ski}, {Sol}, \&
  {Kus}}]{2001A&A...367..809K}
{Katarzy{\'n}ski}, K., {Sol}, H., \& {Kus}, A. 2001, \aap, 367, 809,
  \dodoi{10.1051/0004-6361:20000538}

\bibitem[{{Katz} {et~al.}(2009){Katz}, {Budnik}, \&
  {Waxman}}]{2009JCAP...03..020K}
{Katz}, B., {Budnik}, R., \& {Waxman}, E. 2009, \jcap, 2009, 020,
  \dodoi{10.1088/1475-7516/2009/03/020}

\bibitem[{{Keivani} {et~al.}(2018){Keivani}, {Murase}, {Petropoulou}, {Fox},
  {Cenko}, {Chaty}, {Coleiro}, {DeLaunay}, {Dimitrakoudis}, {Evans}, {Kennea},
  {Marshall}, {Mastichiadis}, {Osborne}, {Santander}, {Tohuvavohu}, \&
  {Turley}}]{2018ApJ...864...84K}
{Keivani}, A., {Murase}, K., {Petropoulou}, M., {et~al.} 2018, \apj, 864, 84,
  \dodoi{10.3847/1538-4357/aad59a}

\bibitem[{{Kelner} \& {Aharonian}(2008)}]{2008PhRvD..78c4013K}
{Kelner}, S.~R., \& {Aharonian}, F.~A. 2008, \prd, 78, 034013,
  \dodoi{10.1103/PhysRevD.78.034013}

\bibitem[{{Komossa}(2008)}]{2008RMxAC..32...86K}
{Komossa}, S. 2008, in Revista Mexicana de Astronomia y Astrofisica Conference
  Series, Vol.~32, Revista Mexicana de Astronomia y Astrofisica Conference
  Series, 86--92.
\newblock \doarXiv{0710.3326}

\bibitem[{Kundu \& Gupta(2014)}]{2014_Kundu}
Kundu, E., \& Gupta, N. 2014, Monthly Notices of the Royal Astronomical
  Society: Letters, 444, L16–L19, \dodoi{10.1093/mnrasl/slu101}

\bibitem[{{Lagage} \& {Cesarsky}(1983)}]{1983A&A...125..249L}
{Lagage}, P.~O., \& {Cesarsky}, C.~J. 1983, \aap, 125, 249

\bibitem[{Liu {et~al.}(2019)Liu, Wang, Xue, Taylor, Wang, Li, \&
  Yan}]{Liu_2019}
Liu, R.-Y., Wang, K., Xue, R., {et~al.} 2019, Physical Review D, 99,
  \dodoi{10.1103/physrevd.99.063008}

\bibitem[{{Mannheim}(1993)}]{1993A&A...269...67M}
{Mannheim}, K. 1993, \aap, 269, 67.
\newblock \doarXiv{astro-ph/9302006}

\bibitem[{{Mannheim}(1995)}]{1995APh.....3..295M}
---. 1995, Astroparticle Physics, 3, 295, \dodoi{10.1016/0927-6505(94)00044-4}

\bibitem[{{Moderski} {et~al.}(2005){Moderski}, {Sikora}, {Coppi}, \&
  {Aharonian}}]{2005MNRAS.363..954M}
{Moderski}, R., {Sikora}, M., {Coppi}, P.~S., \& {Aharonian}, F. 2005, \mnras,
  363, 954, \dodoi{10.1111/j.1365-2966.2005.09494.x}

\bibitem[{{Murase} {et~al.}(2014){Murase}, {Inoue}, \&
  {Dermer}}]{2014PhRvD..90b3007M}
{Murase}, K., {Inoue}, Y., \& {Dermer}, C.~D. 2014, \prd, 90, 023007,
  \dodoi{10.1103/PhysRevD.90.023007}

\bibitem[{{Murase} {et~al.}(2018){Murase}, {Oikonomou}, \&
  {Petropoulou}}]{2018ApJ...865..124M}
{Murase}, K., {Oikonomou}, F., \& {Petropoulou}, M. 2018, \apj, 865, 124,
  \dodoi{10.3847/1538-4357/aada00}

\bibitem[{{Nieppola} {et~al.}(2006){Nieppola}, {Tornikoski}, \&
  {Valtaoja}}]{2006A&A...445..441N}
{Nieppola}, E., {Tornikoski}, M., \& {Valtaoja}, E. 2006, \aap, 445, 441,
  \dodoi{10.1051/0004-6361:20053316}

\bibitem[{Orienti {et~al.}(2012)Orienti, D'Ammando, \&
  Giroletti}]{Orienti_2012}
Orienti, M., D'Ammando, F., \& Giroletti, M. 2012, High resolution radio
  observations of gamma-ray emitting Narrow-Line Seyfert 1s.
\newblock \doarXiv{1205.0402}

\bibitem[{{Osterbrock} \& {Pogge}(1985)}]{Osterbrock_1985}
{Osterbrock}, D.~E., \& {Pogge}, R.~W. 1985, \apj, 297, 166,
  \dodoi{10.1086/163513}

\bibitem[{{Padovani} {et~al.}(2018){Padovani}, {Giommi}, {Resconi}, {Glauch},
  {Arsioli}, {Sahakyan}, \& {Huber}}]{2018MNRAS.480..192P}
{Padovani}, P., {Giommi}, P., {Resconi}, E., {et~al.} 2018, \mnras, 480, 192,
  \dodoi{10.1093/mnras/sty1852}

\bibitem[{{Padovani} {et~al.}(2019){Padovani}, {Oikonomou}, {Petropoulou},
  {Giommi}, \& {Resconi}}]{2019MNRAS.484L.104P}
{Padovani}, P., {Oikonomou}, F., {Petropoulou}, M., {Giommi}, P., \& {Resconi},
  E. 2019, \mnras, 484, L104, \dodoi{10.1093/mnrasl/slz011}

\bibitem[{{Paliya}(2015)}]{2015ApJ...808L..Paliya}
{Paliya}, V.~S. 2015, \apjl, 808, L48, \dodoi{10.1088/2041-8205/808/2/L48}

\bibitem[{Paliya {et~al.}(2018)Paliya, Ajello, Rakshit, Mandal, Stalin, Kaur,
  \& Hartmann}]{Paliya_2018}
Paliya, V.~S., Ajello, M., Rakshit, S., {et~al.} 2018, The Astrophysical
  Journal, 853, L2, \dodoi{10.3847/2041-8213/aaa5ab}

\bibitem[{Paliya {et~al.}(2019)Paliya, Parker, Jiang, Fabian, Brenneman,
  Ajello, \& Hartmann}]{Paliya_2019}
Paliya, V.~S., Parker, M.~L., Jiang, J., {et~al.} 2019, The Astrophysical
  Journal, 872, 169, \dodoi{10.3847/1538-4357/ab01ce}

\bibitem[{Paliya \& Stalin(2016)}]{Paliya_2016}
Paliya, V.~S., \& Stalin, C.~S. 2016, The Astrophysical Journal, 820, 52,
  \dodoi{10.3847/0004-637x/820/1/52}

\bibitem[{{Pierre Auger Collaboration} {et~al.}(2013){Pierre Auger
  Collaboration}, {Abreu}, {Aglietta}, {Ahlers}, {Ahn}, {Albuquerque},
  {Allard}, {Allekotte}, {Allen}, {Allison}, {Almela}, {Alvarez Castillo},
  {Alvarez-Mu{\~n}iz}, {Alves Batista}, {Ambrosio}, {Aminaei}, {Anchordoqui},
  {Andringa}, {Anti{\v{c}}i'c}, {Aramo}, {Arganda}, {Arqueros}, {Asorey},
  {Assis}, {Aublin}, {Ave}, {Avenier}, {Avila}, {Badescu}, {Balzer}, {Barber},
  {Barbosa}, {Bardenet}, {Barroso}, {Baughman}, {B{\"a}uml}, {Baus}, {Beatty},
  {Becker}, {Bell{\'e}toile}, {Bellido}, {BenZvi}, {Berat}, {Bertou},
  {Biermann}, {Billoir}, {Blanco}, {Blanco}, {Bleve}, {Bl{\"u}mer},
  {Boh{\'a}{\v{c}}ov{\'a}}, {Boncioli}, {Bonifazi}, {Bonino}, {Borodai},
  {Brack}, {Brancus}, {Brogueira}, {Brown}, {Bruijn}, {Buchholz}, {Bueno},
  {Buroker}, {Burton}, {Caballero-Mora}, {Caccianiga}, {Caramete}, {Caruso},
  {Castellina}, {Catalano}, {Cataldi}, {Cazon}, {Cester}, {Chauvin}, {Cheng},
  {Chiavassa}, {Chinellato}, {Chirinos Diaz}, {Chudoba}, {Cilmo}, {Clay},
  {Cocciolo}, {Collica}, {Coluccia}, {Concei{\c{c}}{\~a}o}, {Contreras},
  {Cook}, {Cooper}, {Coppens}, {Cordier}, {Coutu}, {Covault}, {Creusot},
  {Criss}, {Cronin}, {Curutiu}, {Dagoret-Campagne}, {Dallier}, {Daniel},
  {Dasso}, {Daumiller}, {Dawson}, {de Almeida}, {De Domenico}, {De Donato}, {de
  Jong}, {De La Vega}, {de Mello Junior}, {de Mello Neto}, {De Mitri}, {de
  Souza}, {de Vries}, {del Peral}, {del R{\'\i}o}, {Deligny}, {Dembinski},
  {Dhital}, {Di Giulio}, {D{\'\i}az Castro}, {Diep}, {Diogo}, {Dobrigkeit},
  {Docters}, {D'Olivo}, {Dong}, {Dorofeev}, {dos Anjos}, {Dova}, {D'Urso},
  {Dutan}, {Ebr}, {Engel}, {Erdmann}, {Escobar}, {Espadanal}, {Etchegoyen},
  {Facal San Luis}, {Falcke}, {Fang}, {Farrar}, {Fauth}, {Fazzini}, {Ferguson},
  {Fick}, {Figueira}, {Filevich}, {Filip{\v{c}}i{\v{c}}}, {Fliescher},
  {Fracchiolla}, {Fraenkel}, {Fratu}, {Fr{\"o}hlich}, {Fuchs}, {Gaior},
  {Gamarra}, {Gambetta}, {Garc{\'\i}a}, {Garcia Roca}, {Garcia-Gamez},
  {Garcia-Pinto}, {Garilli}, {Gascon Bravo}, {Gemmeke}, {Ghia}, {Giller},
  {Gitto}, {Glass}, {Gold}, {Golup}, {Gomez Albarracin}, {G{\'o}mez Berisso},
  {G{\'o}mez Vitale}, {Gon{\c{c}}alves}, {Gonzalez}, {Gookin}, {Gorgi},
  {Gouffon}, {Grashorn}, {Grebe}, {Griffith}, {Grillo}, {Guardincerri},
  {Guarino}, {Guedes}, {Hansen}, {Harari}, {Harrison}, {Harton}, {Haungs},
  {Hebbeker}, {Heck}, {Herve}, {Hill}, {Hojvat}, {Hollon}, {Holmes}, {Homola},
  {H{\"o}randel}, {Horvath}, {Hrabovsk{\'y}}, {Huber}, {Huege}, {Insolia},
  {Ionita}, {Italiano}, {Jansen}, {Jarne}, {Jiraskova}, {Josebachuili},
  {Kadija}, {Kampert}, {Karhan}, {Kasper}, {Katkov}, {K{\'e}gl}, {Keilhauer},
  {Keivani}, {Kelley}, {Kemp}, {Kieckhafer}, {Klages}, {Kleifges},
  {Kleinfeller}, {Knapp}, {Koang}, {Kotera}, {Krohm}, {Kr{\"o}mer},
  {Kruppke-Hansen}, {Kuempel}, {Kulbartz}, {Kunka}, {La Rosa}, {Lachaud},
  {LaHurd}, {Latronico}, {Lauer}, {Lautridou}, {Le Coz}, {Le{\~a}o}, {Lebrun},
  {Lebrun}, {Leigui de Oliveira}, {Letessier-Selvon}, {Lhenry-Yvon}, {Link},
  {L{\'o}pez}, {Lopez Ag{\"u}era}, {Louedec}, {Lozano Bahilo}, {Lu}, {Lucero},
  {Ludwig}, {Lyberis}, {Maccarone}, {Macolino}, {Maldera}, {Maller}, {Mandat},
  {Mantsch}, {Mariazzi}, {Marin}, {Marin}, {Maris}, {Marquez Falcon},
  {Marsella}, {Martello}, {Martin}, {Martinez}, {Mart{\'\i}nez Bravo},
  {Martraire}, {Mas{\'\i}as Meza}, {Mathes}, {Matthews}, {Matthews},
  {Matthiae}, {Maurel}, {Maurizio}, {Mazur}, {Medina-Tanco}, {Melissas},
  {Melo}, {Menichetti}, {Menshikov}, {Mertsch}, {Messina}, {Meurer},
  {Meyhandan}, {Mi'canovi'c}, {Micheletti}, {Minaya}, {Miramonti},
  {Molina-Bueno}, {Mollerach}, {Monasor}, {Monnier Ragaigne}, {Montanet},
  {Morales}, {Morello}, {Moreno}, {Moreno}, {Mostaf{\'a}}, {Moura}, {Muller},
  {M{\"u}ller}, {M{\"u}nchmeyer}, {Mussa}, {Navarra}, {Navarro}, {Navas},
  {Necesal}, {Nellen}, {Nelles}, {Neuser}, {Nhung}, {Niechciol}, {Niemietz},
  {Nierstenhoefer}, {Nitz}, {Nosek}, {No{\v{z}}ka}, {Oehlschl{\"a}ger},
  {Olinto}, {Ortiz}, {Pacheco}, {Pakk Selmi-Dei}, {Palatka}, {Pallotta},
  {Palmieri}, {Parente}, {Parizot}, {Parra}, {Pastor}, {Paul}, {Pech},
  {Pe{\c{k}}ala}, {Pelayo}, {Pepe}, {Perrone}, {Pesce}, {Petermann}, {Petrera},
  {Petrolini}, {Petrov}, {Pfendner}, {Piegaia}, {Pierog}, {Pieroni}, {Pimenta},
  {Pirronello}, {Platino}, {Plum}, {Ponce}, {Pontz}, {Porcelli}, {Privitera},
  {Prouza}, {Quel}, {Querchfeld}, {Rautenberg}, {Ravel}, {Ravignani}, {Revenu},
  {Ridky}, {Riggi}, {Risse}, {Ristori}, {Rivera}, {Rizi}, {Roberts}, {Rodrigues
  de Carvalho}, {Rodriguez}, {Rodriguez Cabo}, {Rodriguez Martino}, {Rodriguez
  Rojo}, {Rodr{\'\i}guez-Fr{\'\i}as}, {Ros}, {Rosado}, {Rossler}, {Roth},
  {Rouill{\'e}-d'Orfeuil}, {Roulet}, {Rovero}, {R{\"u}hle}, {Saftoiu},
  {Salamida}, {Salazar}, {Salesa Greus}, {Salina}, {S{\'a}nchez}, {Santo},
  {Santos}, {Santos}, {Sarazin}, {Sarkar}, {Sarkar}, {Sato}, {Scharf},
  {Scherini}, {Schieler}, {Schiffer}, {Schmidt}, {Scholten}, {Schoorlemmer},
  {Schovancova}, {Schov{\'a}nek}, {Schr{\"o}der}, {Schuster}, {Sciutto},
  {Scuderi}, {Segreto}, {Settimo}, {Shadkam}, {Shellard}, {Sidelnik}, {Sigl},
  {Silva Lopez}, {Sima}, {'Smia{\l}kowski}, {{\v{S}}m{\'\i}da}, {Snow},
  {Sommers}, {Sorokin}, {Spinka}, {Squartini}, {Srivastava}, {Stanic},
  {Stapleton}, {Stasielak}, {Stephan}, {Stutz}, {Suarez}, {Suomij{\"a}rvi},
  {Supanitsky}, {{\v{S}}u{\v{s}}a}, {Sutherland}, {Swain}, {Szadkowski},
  {Szuba}, {Tapia}, {Tartare}, {Ta{\c{s}}c{\u{a}}u}, {Tcaciuc}, {Thao},
  {Thomas}, {Tiffenberg}, {Timmermans}, {Tkaczyk}, {Todero Peixoto}, {Toma},
  {Tomankova}, {Tom{\'e}}, {Tonachini}, {Torralba Elipe}, {Travnicek},
  {Tridapalli}, {Tristram}, {Trovato}, {Tueros}, {Ulrich}, {Unger}, {Urban},
  {Vald{\'e}s Galicia}, {Vali{\~n}o}, {Valore}, {van Aar}, {van den Berg}, {van
  Velzen}, {van Vliet}, {Varela}, {Vargas C{\'a}rdenas}, {V{\'a}zquez},
  {V{\'a}zquez}, {Veberi{\v{c}}}, {Verzi}, {Vicha}, {Videla}, {Villase{\~n}or},
  {Wahlberg}, {Wahrlich}, {Wainberg}, {Walz}, {Watson}, {Weber}, {Weidenhaupt},
  {Weindl}, {Werner}, {Westerhoff}, {Whelan}, {Widom}, {Wieczorek}, {Wiencke},
  {Wilczy{\'n}ska}, {Wilczy{\'n}ski}, {Will}, {Williams}, {Winchen}, {Wommer},
  {Wundheiler}, {Yamamoto}, {Yapici}, {Younk}, {Yuan}, {Yushkov}, {Zamorano
  Garcia}, {Zas}, {Zavrtanik}, {Zavrtanik}, {Zaw}, {Zepeda}, {Zhou}, {Zhu},
  {Zimbres Silva}, \& {Ziolkowski}}]{Pierre_2013}
{Pierre Auger Collaboration}, {Abreu}, P., {Aglietta}, M., {et~al.} 2013,
  \apjl, 762, L13, \dodoi{10.1088/2041-8205/762/1/L13}

\bibitem[{{Reimer} {et~al.}(2019){Reimer}, {B{\"o}ttcher}, \&
  {Buson}}]{2019ApJ...881...46R}
{Reimer}, A., {B{\"o}ttcher}, M., \& {Buson}, S. 2019, \apj, 881, 46,
  \dodoi{10.3847/1538-4357/ab2bff}

\bibitem[{{Rodrigues} {et~al.}(2019){Rodrigues}, {Gao}, {Fedynitch},
  {Palladino}, \& {Winter}}]{2019ApJ...874L..29R}
{Rodrigues}, X., {Gao}, S., {Fedynitch}, A., {Palladino}, A., \& {Winter}, W.
  2019, \apjl, 874, L29, \dodoi{10.3847/2041-8213/ab1267}

\bibitem[{{Sahakyan}(2018)}]{2018ApJ...866..109S}
{Sahakyan}, N. 2018, \apj, 866, 109, \dodoi{10.3847/1538-4357/aadade}

\bibitem[{{Sambruna} {et~al.}(1996){Sambruna}, {Maraschi}, \&
  {Urry}}]{1996ApJ...463..444S}
{Sambruna}, R.~M., {Maraschi}, L., \& {Urry}, C.~M. 1996, \apj, 463, 444,
  \dodoi{10.1086/177260}

\bibitem[{Schr\"oder {et~al.}(2019)}]{Schroder_2019}
Schr\"oder, F.~G., {et~al.} 2019, Bull. Am. Astron. Soc., 51, 131.
\newblock \doarXiv{1903.07713}

\bibitem[{{Sikora}(2011)}]{2011IAUS..275...59S}
{Sikora}, M. 2011, in Jets at All Scales, ed. G.~E. {Romero}, R.~A. {Sunyaev},
  \& T.~{Belloni}, Vol. 275, 59--67, \dodoi{10.1017/S1743921310015644}

\bibitem[{{Stecker} \& {Salamon}(1996)}]{1996SSRv...75..341S}
{Stecker}, F.~W., \& {Salamon}, M.~H. 1996, \ssr, 75, 341,
  \dodoi{10.1007/BF00195044}

\bibitem[{{Tavecchio} \& {Ghisellini}(2008)}]{2008MNRAS.386..945T}
{Tavecchio}, F., \& {Ghisellini}, G. 2008, \mnras, 386, 945,
  \dodoi{10.1111/j.1365-2966.2008.13072.x}

\bibitem[{{Urry} \& {Padovani}(1995)}]{1995PASP..107..803U}
{Urry}, C.~M., \& {Padovani}, P. 1995, \pasp, 107, 803, \dodoi{10.1086/133630}

\bibitem[{{von Montigny} {et~al.}(1995){von Montigny}, {Bertsch}, {Chiang},
  {Dingus}, {Esposito}, {Fichtel}, {Fierro}, {Hartman}, {Hunter}, {Kanbach},
  {Kniffen}, {Lin}, {Mattox}, {Mayer-Hasselwander}, {Michelson}, {Nolan},
  {Radecke}, {Schneid}, {Sreekumar}, {Thompson}, \&
  {Willis}}]{1995ApJ...440..525V}
{von Montigny}, C., {Bertsch}, D.~L., {Chiang}, J., {et~al.} 1995, \apj, 440,
  525, \dodoi{10.1086/175294}

\bibitem[{Wang {et~al.}(2020)Wang, Zhang, Sun, \& Liang}]{2020_Zhenjie}
Wang, Z.-J., Zhang, J., Sun, X.-N., \& Liang, E.-W. 2020, The Astrophysical
  Journal, 893, 41, \dodoi{10.3847/1538-4357/ab7d35}

\bibitem[{Wang {et~al.}(2022)Wang, Liu, Petropoulou, Oikonomou, Xue, \&
  Wang}]{Wang_2022}
Wang, Z.-R., Liu, R.-Y., Petropoulou, M., {et~al.} 2022, Physical Review D,
  105, \dodoi{10.1103/physrevd.105.023005}

\bibitem[{{Xue} {et~al.}(2019){Xue}, {Liu}, {Petropoulou}, {Oikonomou}, {Wang},
  {Wang}, \& {Wang}}]{Xue_2019}
{Xue}, R., {Liu}, R.-Y., {Petropoulou}, M., {et~al.} 2019, \apj, 886, 23,
  \dodoi{10.3847/1538-4357/ab4b44}

\bibitem[{Zhang {et~al.}(2020)Zhang, Zhang, Gan, Yi, Wang, \&
  Liang}]{Zhang_2020}
Zhang, J., Zhang, H.-M., Gan, Y.-Y., {et~al.} 2020, The Astrophysical Journal,
  899, 2, \dodoi{10.3847/1538-4357/aba2cd}

\end{thebibliography}
\bibliographystyle{aasjournal}


\newpage
\appendix
\renewcommand{\appendixname}{Appendix~\Alph{section}}
Here we present the plots of MCMC fitting under the leptonic model, the proton synchrotron model, and the photohadronic model, respectively.

\begin{figure}[ht!]
\plotone{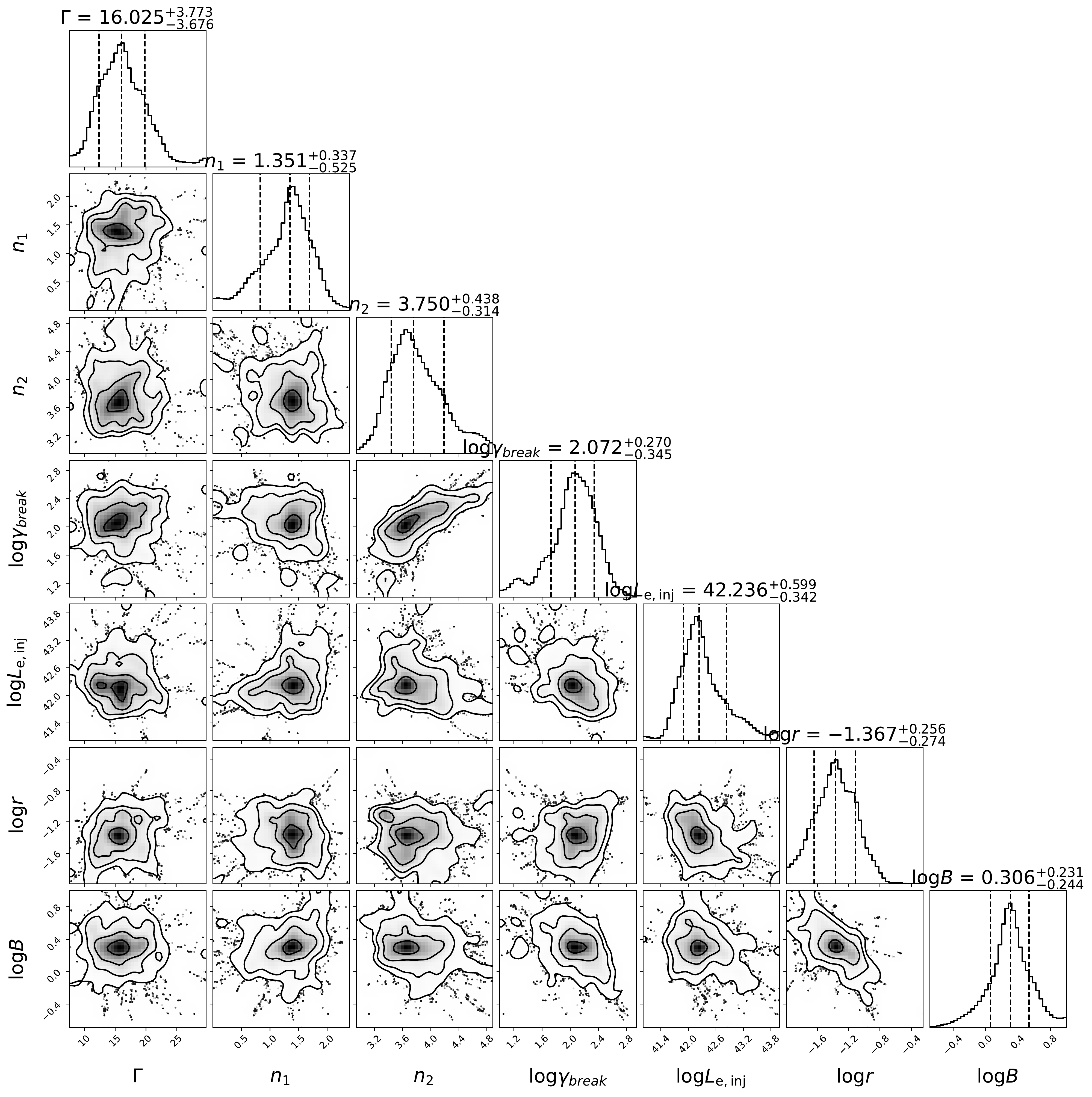}
\caption{The result of MCMC under the leptonic model\label{fig:general}, the three dotted lines represent the uncertainties based on the 16th, 50th, and 84th percentiles of the samples in the marginalized distributions.}
\end{figure}

\begin{figure}[ht!]
\plotone{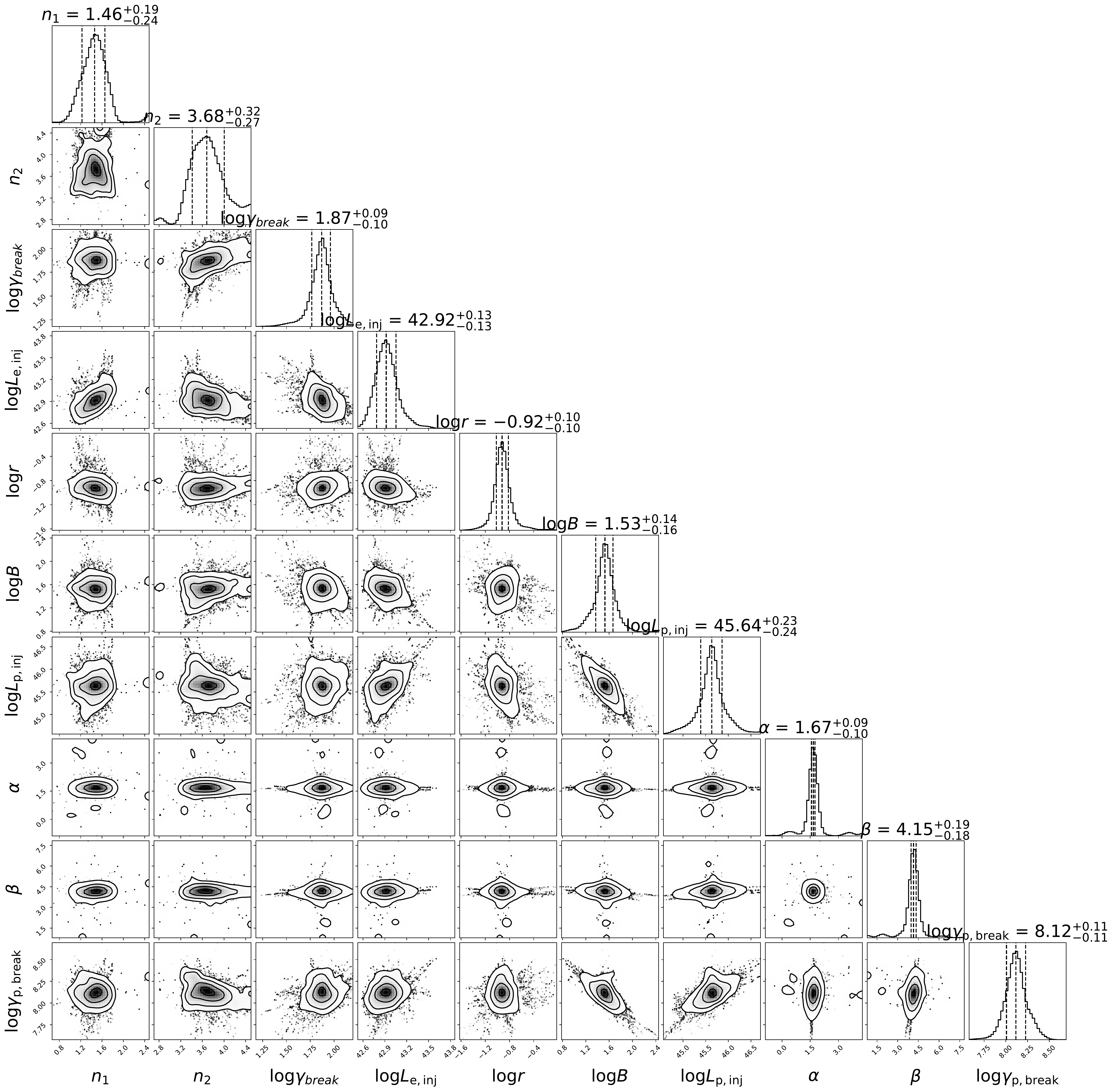}
\caption{The result of MCMC under the proton synchrotron model\label{fig:general}, the three dotted lines represent the uncertainties based on the 16th, 50th, and 84th percentiles of the samples in the marginalized distributions.}
\end{figure}

\begin{figure}[ht!]
\plotone{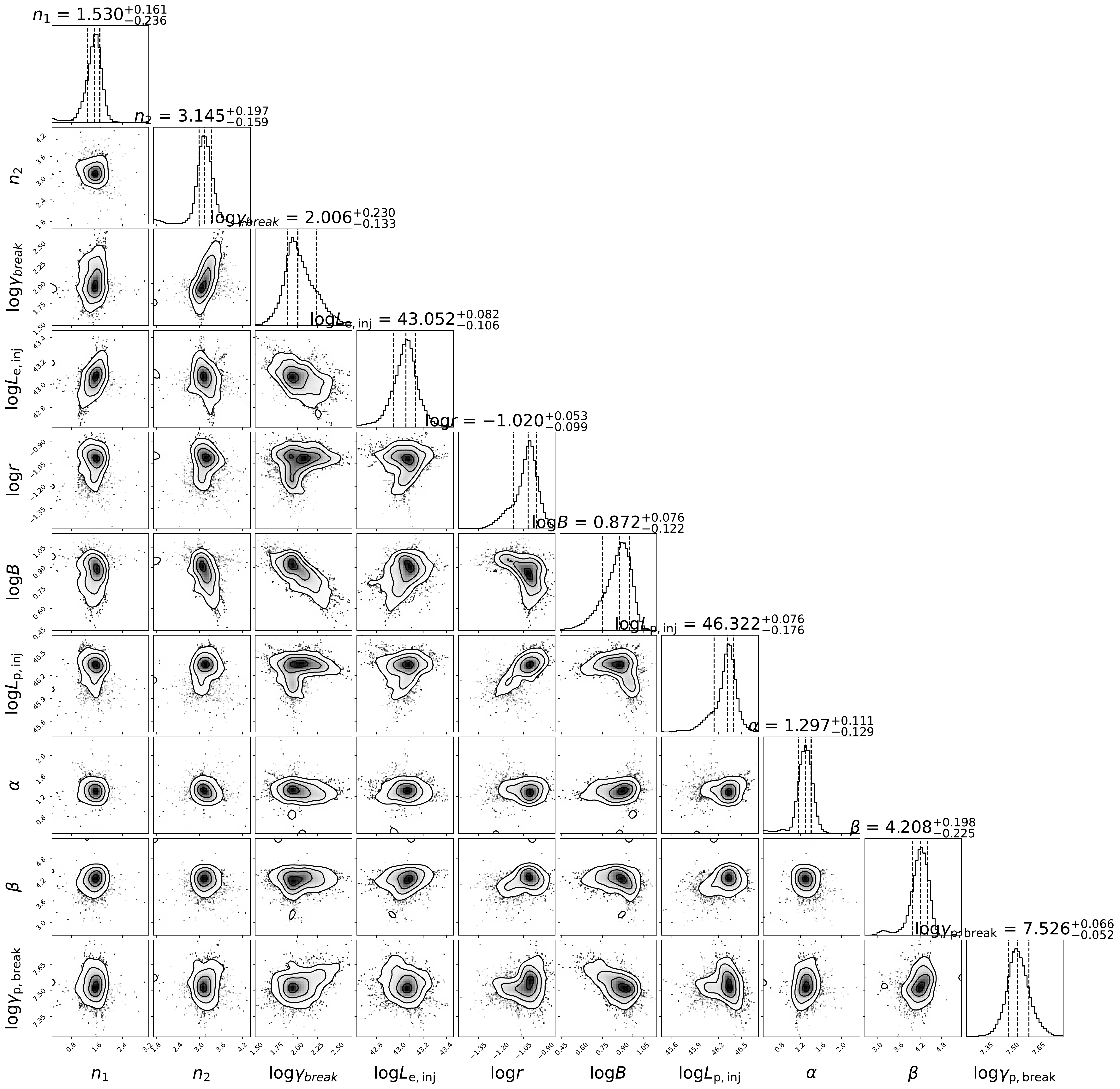}
\caption{The result of MCMC under the photohadronic model \label{fig:general}, the three dotted lines represent the uncertainties based on the 16th, 50th, and 84th percentiles of the samples in the marginalized distributions.}
\end{figure}

\end{document}